\documentclass[aps,prd,preprint,groupedaddress, amsfonts,amssymb,amsmath, floatfix]{revtex4}
\usepackage{natbib}
\usepackage{bm}
\usepackage{graphicx}
\usepackage{supertabular}
\usepackage{rotating}
\begin{document}

\title{Volatilities That Change with Time:
The Temporal Behavior of the Distribution of Stock-Market Prices}  

\author{A.~D.~Speliotopoulos}

\email{achilles@cal.berkeley.edu}

\affiliation{
Department of Mathematics,
Golden Gate University,
San Francisco, CA 94105
}

\date{July 29, 2010}

\begin{abstract}

While the use of volatilities is pervasive throughout finance, our ability
to determine the instantaneous volatility of stocks is nascent. Here, 
we present a method for measuring the temporal behavior of
stocks, and show that stock prices for 24 DJIA stocks follow a stochastic
process that describes an efficiently priced stock while using a
volatility that changes deterministically with time. We find that the
often observed, abnormally large kurtoses are due to temporal
variations in the volatility. Our method can resolve changes in
volatility and drift of the stocks as fast as a single day using
daily close prices. 
 
\vskip 24pt

\noindent{keywords: spectral analysis, noise
  reduction, Rademacher distribution} 

\end{abstract}


\maketitle

\section{Introduction}
\label{sec: 0}
In this paper, we study the temporal behavior of the distribution of
stock prices for 24 stocks in the Dow Jones Industrial Average
(DJIA). This is done using a new method of measuring changes in the
volatility and drifts of stocks with time. When this method is
applied to time-series constructed from the daily close of stocks,
changes as fast as one day can be seen in both.  Given that it is not
possible to accurately \textit{measure} (as oppose to
\textit{predict}) intraday changes in the volatility using only
daily-close data, for two of the 24 stocks we have been able to
reach the maximum resolution (known as the Nyquist criteria) of one
day in the rate that the volatility can change, while for the great
majority of the remaining stocks, we have come within one day of this
maximum.  We believe that this method can measure changes in the
volatility and drift that occur during the trading day as well if
intraday price data is used. But even with only daily-close data, we
have been extraordinarily successful at determining the temporal
behavior of stocks in general, and of the volatility in particular,
and in the process, we have furthered our understanding of the
behavior of stock prices as a whole.  

We find that the stock prices of these 24 stocks can be well
described by a  stochastic process for which the volatility changes
\textit{deterministically} with time. On the one hand, this is a process
where the yield at any one time is not correlated with the yield at
any other time; the process thus describes an efficiently priced
stock. On the other hand, this is a process where the predicted
kurtosis agrees with the sample kurtosis of the stock; the process
thus also provides a solution to the long standing problem of
explaining how an efficiently priced stock can have a kurtosis that is
so different from what is expected for a Gaussian
distribution. Indeed, we find that abnormally large kurtoses are  
due solely to changes in the volatility of the stock with time. When
this temporal behavior is accounted for in the daily yield, the
kurtosis reduces dramatically in value, and now agrees well with model
predictions. This finding is in agreement with Rosenberg's (1972) 
observation that the kurtosis for nonstationary random variables 
is larger than than the kurtosis of individual random
variables. We have also determined changes in the volatility of these
stocks, and for three of the 24 stocks, variations of as fast as 
one day can be seen. For another 16 stocks, this temporal resolution
was two days or less, and for only five of the 24 stocks is this resolution
longer than 2.5 days.  

The behavior of the drifts for all 24 stocks can also be determined using
this method, and with the same resolution as their volatility.  We
find that the drift for the majority of the stocks is positive; these
drifts thus tend to augment the increase of the stock 
price caused by the random-walk nature of the stochastic process. This
finding is not surprising, nor is it surprising that we find that the
drift is much smaller than the volatility for all 24 stocks. What is
surprising is that for three of the 24 stocks the drift is uniformly
\textit{negative}. For these stocks, the drift tends not to increase
the stock price, but to depress it. That the stock price for these
three stocks increase at all is because this drift is
much smaller in the magnitude than the volatility. Over the short
term, growth in the prices of these stocks\textemdash as they are for
all 24 stocks\textemdash is due to a random walk, and thus driven
more by the volatility than the drift. Indeed, this is the only reason
that the prices of these stocks increase with time.  

Finally, the distribution of the stock prices for the 24 DJIA stocks
has been determined. When the temporal variation in the volatility is
corrected for in the daily yield, we find that the resultant
distribution for all but four of the stocks is described by a Rademacher
distribution with the probability that the yield increases on any one
day being 1/2. For the four other stocks, the distribution is
described by a generalized Rademacher distribution with the
probability that the yield increases on any one day being slightly
greater than the probability that it decreases.    
 
\section{Background, Previous Work, and a Summary of the Approach}

In 2005, 403.8 billion shares were traded on the New York Stock 
Exchange (NYSE) with a total value of \$14.1 trillion dollars (see
NYSE). During the same period, 468 million contracts were written on
the Chicago Board Options Exchange (CBOE) with a total notional value
of \$12 trillion dollars. At the NYSE, traders, investors, and
speculators\textemdash big and small\textemdash place bets on the
movement of stock prices, whether up or down. Profits are made, or
losses are reconciled, based on the changing price of the stock. As
such, great effort is made to predict the movements of stock prices in
the future, and thus much attention\textemdash with attending
analysis\textemdash is focused on the price of stocks.  

In the CBOE, traders, investors, and speculators write or enter into
contacts to purchase or sell a predetermined amount of stocks at a set time in
the future. Profits here are made, or losses reconciled, based on the
degree of risk that the movement of the stock will be down when
expected to be up, or up when expected to be down. Here, it is not so
much the price of the stock that matters. It is the amount of
volatility in the stock, and predicting how stock prices may move in
the future is much less important. Indeed, the pricing of
options\textemdash through the Black-Scholes equation and its
variants\textemdash is based on the argument that it is \textit{not}
possible to predict how the price of stocks will change in the
future. In this pricing, it is taken for granted that the markets are
efficient, and that earning returns which are in excess of the
risk-free interest rate is not possible. All is random, and the
increase in stock prices seen is due to a simple random walk with a
(small) drift. Great interest is thus paid in modeling the
\textit{distribution} of stock prices, and the application of these
models to the pricing of options and derivatives.  

Given the \$26.1 trillion dollars in trades and contracts in the NYSE
and CBOE in 2005, it is not surprising that much effort has been
expended in determining the properties of the stock market. Given the
precipitous drop in stock market prices in October of 2008\textemdash
\textit{which occurred over period of days}\textemdash accurate
determination of how these properties change with time has become
even more important. Since the work by Bachelier (1900) at the turn of
the 20th century, a great deal of these efforts have been focused on
determining the distribution of the daily yields of stock prices
(Osborne 1959a and Osborne 1959b).  Inherent in this determination
is determining the volatility of the distribution. Use of this
volatility is now pervasive in modern finance, and is a critical
ingredient in such endeavors as the pricing of options, the general
assessment of risk and the determination the value of assets 
at risk, and the construction of optimal portfolios. That this effort
continues today is indicative of the difficulty in determining this
distribution, its importance in modern finance, and the
financial impact that its determination can have. 

While Bachelier (1990) characterized the distribution as a
random walk with the prices of the stock having a given drift and a
constant volatility, it has been known since the detail analysis of the
behavior of stock prices by Fama (1965) that the distribution
of daily yields is only approximately Gaussian; the distribution
calculated by Fama\textemdash which does not take into account
variations in the volatility with time\textemdash has a fatter tail
than expected for a Gaussian distribution. Indeed, it is typically
found that the kurtosis can be as high as 100, while by comparison the
kurtosis of a Gaussian distribution is only three. This discrepancy between
the distribution of daily yields as they are traditionally calculated
and the Gaussian distribution, while seemingly an inconsequential detail,
nonetheless has wide-ranging consequences.  

Mathematics tells us that if the distribution of daily yields of a
stock is a Gaussian distribution, then the daily yield on any one day
cannot depend on the daily yield on any other. This is the Central
Limit Theorem (CLT), and it is embodied in a number of ways\textemdash the
various forms of the Efficient Market Hypothesis (EMH) (see Fama 1970 and
Fama and French 1988), and the no-arbitrage condition\textemdash in
modern finance. This lack of predictability is one of the underlying
assumptions used in the pricing of derivatives. Mathematics says we
can also turn the statement of the CLT around, however. Namely, if the
daily yield on any one day does not depend on the daily yield on any
other, then the distribution of daily yields must \textit{necessarily}
be a Gaussian distribution as long as the number of days used in its
determination is large enough, and as long as the distribution is well
behaved.

In the face of this mathematical result, there are two
possibilities. The first possibility is that 
the distribution of daily yields for stocks is not Gaussian. The daily
yield on one day does depend on the daily yield on some other day, and
it is possible, in principle, to predict future stock prices by looking at
historical prices. The second possibility is that the EMH nevertheless
holds, and there are good, albeit unknown, reasons for the
unexpectedly large kurtosis. The situation is further muddied when
the autocorrelations of the daily yield of stocks are calculated. It
is  well known from these calculations that the value of the daily
yield on different days are uncorrelated with each other, and we have
seen this behavior for the stocks studied here as well. This
independence extends also to other asset classes, as shown by Kendall
(1953). 

There have been numerous attempts at using other distributions\textemdash the
Levy and its generalization, the Pareto, proposed by Mandelbrot (1963),
the Student t-Distribution proposed by Blattberg and Gonedes (1974),
and the discrete mixture of Gaussian distributions model proposed by
Kon (1984)\textemdash to describe the distribution of stock prices
(see T\"oyli, Sysi-Aho, and Kaski 2004 for an overview and 
assessment). These attempts are based on the belief that the second of
the two possibilities holds, and that the reason for the overly large
kurtosis is because the distribution used to describe the stock was not
correct. As such, for these distributions the daily yield on
any one day also does not depend on the daily yield on any other day,
and the consequences of the CLT is instead evaded in various ways. These
approaches have had various degrees of success. For example, while the
Pareto distribution does have a fatter tail than the Gaussian
distribution and has a kurtosis that can agree with 
observations, all moments with order greater than an integer
$k$\textemdash which determines the power-law behavior of the
distribution\textemdash is ill defined; in this way, the 
distribution is not well behaved, and thus does not fall within the
class of distributions for which the CLT is applicable. For the Levy
distribution, the volatility itself (as well as all higher moments) is
ill-defined, requiring the truncation of the distribution to price
options using this model, as described in Kleinert (2002). The
Student-t Distribution differs significantly from the Gaussian
distribution only when the number of data points are small (thereby
evading the CLT), which begs the question of what happens when this
distribution is applied to time-series with more than, say, 200 terms
in it. Kon's model is a of mixture of Gaussian distributions, and thus
the moments of his distribution are all finite. However, while the
model is effective at describing the large kurtosis of stocks, it is
nonetheless an empirical model; the origin of the discrete mixture is
not known, and the parameters used in its construction are determined
only after the model is fitted to the stock data. 

Our approach is also based on the belief that the second of the two
possibilities hold.  But unlike the previous attempts at describing the
distribution of stock prices mentioned above, we find that the
underlying reason for the overly large kurtosis is because time
variations in the distribution of stocks have not been properly taken
account of. As observed by Rosenberg (1972), it is often assumed that the  
distribution  of stock prices being analyzed does not change during
the period of interest. This assumption was certainly made for all the
models described above. In contrast to these approaches, we will take
time variations in the distribution explicitly into account. Doing so
results in a distribution that can both explain the abnormally
kurtosis, and still have the property that the yield on any one day is
not correlated with the yield on any other. In the process, we will also
be able to determine, for the first time, how the volatility and the
drift changes instantaneously with time. 
  
That the volatility of stocks changes with time is not a new
observation. This behavior has been known since at least
the work by Osborne (1962) (see also Lo 1988), and analyzed
explicitly by Rosenberg (1972). Indications of this have been reported
by many others since then (Ball and Torous 1985, French and Roll
1986, Conrad and Kaul 1988, Andersen and Bollerslev 1997,
Kullmann, T\"oyli, Kertesz, Kanto, and Kaski 1999, Nawroth and
Peinke 2006). Much effort has since been made to determine how this 
volatility\textemdash and thus necessarily how the distribution\textemdash
changes with time, with the main focus of this effort on extending
the usual random walk description of stock market prices. This has
lead to the introduction of the jump-diffusion model proposed by
Merton (1976), where discrete, random jumps in the prices of a stock
in time are incorporated in continuous stochastic processes, and to
stochastic volatility models developed over a number of years by
Praetz (1972), Christie (1982), Hull and White (1987), Scott (1987)
and Heston (1993) where the volatility itself is modeled as a
stochastic process with its own drift and volatility  (see also Muzy,
Delour, and Bacry 2000 for a multifractal-inspired, stochastic
volatility model). However, as it was pointed out by Hull and White
(1987), methods for directly \textit{measuring} time-varying
volatilities were not, at the time, known.    

This inability to directly measure variations in the volatility has
greatly constrained efforts in studying how the volatility of real
market data varies with time. To a great extent, it has also driven
the development of stochastic volatility models. By 
characterizing the volatility as a stochastic process, a time varying
volatility can be modeled using a comparatively simple choice of a
constant drift and a constant volatility for the process. Even then, however,
parameters in stochastic volatility models are determined not by a
direct analysis of the daily yields of stock prices, but are instead
determined indirectly. Namely, the price of an option for a stock is
calculated for the process in terms of a set of model parameters, and
these parameters are then set by adjusting their values until the
calculated price agrees with the market price of the option. 

The inherent difficulty in determining from market data how a
distribution changes with time is described in Boyle and
Anathanarayanan (1977), and is straightforward to understand. To
determine the distribution of a stock, a collection of stock prices is
required; the larger the collection, the better. Since stock 
prices change sequentially in time, this collection has to be done
over a period of time, and because a relatively large collection is
needed, this period must be correspondingly long. For example, most
distributions are calculated using the daily close of a stock, if for
no other reason then because these prices are readily available 
in the public domain. If the collection of prices used is as large as
500 daily closes, then the stock prices in this collection must span
a period of nearly two years; Fama (1965), for example, used stock
prices that span a period of up to six years in his analysis. Using a
collection of 500 stock prices to  determine the distribution of the
stock through standard methods means 
that one is tacitly assuming that the price of the stock two years ago belongs
to the same distribution\textemdash\textit{with the same
  volatility}\textemdash as the price of the stock today. This strains
credibility, especially given the rapid movements in the markets during
the last quarter of 2008. While it is possible to calculate the
distribution with a fewer number of stock prices and thus shorten the
period of time over which they are collected, statistical errors
inherent in determining the distribution are proportional to
$1/\sqrt{N}$, where $N$ are the number of data points in the sample,
and will thus be correspondingly larger. At one point, the period of
time would be so short that we would not be able to say whether
distribution is Gaussian or not. Indeed, we only have to look at the
extreme case where the period is so short that there are only three
stock prices collected over three days, resulting in only two daily
yields to determine the whole distribution; this clearly cannot be
done with any certainty whatsoever! (This inherent difficulty has lead
to the development of other approaches to calculating volatility such
as those found in Ball and Torous 1984, Parkinson 1980, and Longin
2005) where the number of daily close needed is reduced.)

Mathematics does not require that the distribution remains
constant. The general theory of stochastic processes allows for
volatilities that change with time. In fact, we will show below that
even though the volatility of a Gaussian distribution may change with
time, the daily yield on any one day need not depend on the daily
yield on any previous day; the EMH still holds for this case. Instead,
what has been lacking up to now is a method for \textit{calculating}
the statistical properties of a stock when the volatility changes with
time. This we have been able to do.

Our approach is based on the observation that when the
volatility depends solely on time, we can remove the time dependence
of the distribution by dividing the daily yield by the
volatility. This standardizes the daily yield, and a Gaussian
distribution with a time-varying volatility is mapped into a Gaussian
distribution with unit volatility. The volatility of this distribution
is clearly constant, and  thus the standardized daily yields all
belong to the \textit{same} distribution. The inherent difficulty in
determining a distribution that changes with time mentioned above
is thus circumvented. Indeed, large collections of stock prices are now a
benefit\textemdash they result in smaller standard errors\textemdash
and not a detriment. That the volatility of the mapped distribution is
\textit{known} then allows us to determine how the volatility of the
original daily yield changes with time. In addition, it is readily
apparent from our analysis that the distribution of standardized
yields is equivalent to a special case of the binomial distribution, and
this observation allows us to extract easily the temporal behavior of
the drift of the yield as well. 

This approach is straightforward, and at its heart resembles the
process one goes through in using a table of values for the cumulative
standardized Gaussian distribution: The random variable at hand is
scaled with its volatility to get the standardized Gaussian
distribution with unit volatility. The difference is that in our case
the volatility is not know a prior\'i; it must be determined. This is
done using a combination of statistical methods, Fourier analysis, and
signal processing techniques. While prevalent in other fields, many of these
techniques are not commonly found in the finance or business literature, and it
would be easy to become too involved with the mathematics while
neglecting the finance when presenting our results. To avoid doing so,
we will focus on finance in the main body of the paper, and when
our model of stock prices is constructed, it will be motivated by, and
justified with, an analysis of the time-series of the stocks at
hand. Importantly, a validation of each step taken will  made. Only
enough of the underlying mathematical analysis needed to explain the
essential ideas behind our approach will be presented in the main body
of the paper; we will refer the reader to the appendices for many of
the details. Our analysis will be applied explicitly to Coca Cola
stock in this paper to demonstrate the underlying ideas behind the
approach. This stock is chosen out of the 24 because for our purposes
its underlying behavior is representative of all the others. Analysis
of the other 23 stocks studied here follow in much the same way, and
we will only present a summary of the results for them, along with
graphs of the volatilities for all 24 stocks as a function of trading day.  
  
\section{Model Validation and Our Choice of Stocks}
\label{sec: 1} 

It would not be an exaggeration to say that the only characterizations
of a stock that is not model dependent is the price per share that it
was sold at, the day and time it was sold, and the total number of
shares of the stock that was sold over a given period of time. These
are the only characterizations that are objective and verifiable,
and for whom all can agree on how they are obtained. The distribution of
the daily yields of a stock certainly is not, and herein lies the
problem: How should any model of stock prices be validated?  

To see how difficult the problem of validation is to resolve (this
issue was explicitly studied by Magdon-Ismail and Abu-Mostafa 1998
for volatility models), consider the volatility of the 24 stocks
considered here. As we will calculate the volatility of these stocks,
it would seem that a comparison of the volatilities we obtain here
with the volatilities calculated using any one of the many other
approaches in the literature would be an effective way of assessing
the validity of  our model. However, irrespective of the approach
taken to make this calculation, assumptions about the behavior of the
stock will have already been made. The 
historical volatility, for example, uses a moving average to calculate
the volatility on any given trading day. It implicitly 
assumes that the volatility does not change significantly over the
window of time used when calculating the average, and thus
cannot effectively measure changes in the volatility that occur within
this window. The implicit volatility, developed over a series of
papers by Latan\'e and Rendleman (1976), Schmalensee and Trippi
(1978), and Beckers (1981), can measure instantaneous changes in the
volatility, but it is calculated by inverting the Black-Scholes (or
any other) equation for pricing options, and thus implicitly assumes
that the particular pricing equation used accurately prices the option
at any given time. Autoregression approaches to calculating the
volatility\textemdash such as the exponentially weighted moving
average (EWMA), the autoregregressive conditional heteroskedasticity
(ARCH) proposed by Engle (1982), the generalized autoregregressive
conditional heteroskedasticity (GARCH) proposed by Bollerslev (1986),
and a new approach that combines autoregressive and Fourier (spectral)
analysis techniques proposed by Bollerslev and Wright
(2001)\textemdash are designed more to \textit{manage} volatilities
that change with time than to \textit{characterize} them. They depend
on one or more parameters that must subsequently be set using some
property of a stock, and are not designed to \textit{determine} how
the volatility changes. Stochastic volatility models explicitly
consider volatilities that change (randomly) with time, but to
determine how this volatility changes, the approach adjusts the
parameters that determine the volatility until the predicted option
prices agree with market prices (see Lamoureux and Lastrapes 1993
for a test of this approach). Using a comparison of volatilities to
validate models is therefore more a comparison of the underlying
models of the market or methods of calculation than it is of the
volatilities themselves. Indeed, the question of which approach to
calculating the volatility is the better one is one that has been
address many times over the years (see Day and Lewis 1992, Canina
and Figlewski 1993, Jorion 1995, Figlewski 1997, Andersen and
Bollerslev 1998, Chong, Ahmad, and Abdullah 1999, Szakmary, Ors,
Kim, and Davidson II 2003, and McMillan and Speight 2004),
apparently without consensus.  

This difficulty in validating models is particularly inopportune
here. While many of the techniques we have used in this paper has been
long used in other fields, our approach in this paper is novel, and
have not been used in the analysis of stock market prices before. We
therefore take a particularly stringent approach to validating our
model. First, the model must be able to explain the observed
properties of the stocks. This we accomplish by
construction. Properties of the stock price are presented
\textit{first}, and the model is then constructed explicitly to
describe them. Second, the model must be self-consistent, and must be
able to predict some property the stock price, which can subsequently be
verified. All models of stock market prices make a certain set of
underlying assumptions about properties of the price; these 
assumptions have consequences. These consequences can in turn be used
to predict properties of the stock price that can then be used to
validate it. For our model, the distribution of standardized daily
yields is described by a Rademacher distribution or its
generalization. This distribution gives specific values for 
the population skewness and kurtosis, and they provide a
simple and statistically meaningful approach to validating our
model. Specifically, we calculate the sample skewness and
kurtosis from each stock's time-series. We then compare this
sample skewness and kurtosis to the population skewness and kurtosis
predicted by our model. If the sample skewness and kurtosis agree with
the population skewness and kurtosis of our model at the 95\%
confidence level (CL), we assert that our model is valid. In fact, we
find that this agreement holds for all 24 stocks considered here, and
it does so over the whole of the  time period spanned by their time
series.  Indeed, for a number of the stocks, \textit{this period spans
  over 80 years}.

It is because of this operational approach to validating our model that
we chose to analyzed stocks from the DJIA. First, all the stocks in
the DJIA are large caps, and have a large daily trading 
volume; they are precisely the type of stocks for which we expect
the market to be efficient. They are in this way similar, and
we would expect they can be described by the same type of model. Second,
each of these companies has been publicly traded for a number of years. We
therefore have access to a large collection of daily close prices for
these stocks with which to construct their time-series. These time
series, for example, range in time from as short as 5,090 trading days for
Citigroup, to as long as 21,527 trading days for Exxon-Mobil. The
availability of a large sample of daily close is particularly
important as we will be numerically assessing the validity of each
step in the construction of our model. With such large collections of
stock prices, standard errors in our calculation can be as small
as 0.7\%, and as such, we are able to say with a great deal of
certainty whether or not our approach is self-consistent. Third, the
24 chosen were the simplest, in terms of their ownership, of the 30
stocks listed in the DJIA. The six DJIA not chosen were recently
involved in mergers or acquisitions, which introduces unwanted
complications; an assessment of the temporal of these stock prices
may not be as clear cut as the 24 stocks considered here. 

A detailed description of how the time-series are constructed is given
in Appendix $\ref{sec: A1}$, where any particularities in the analysis
of stocks are listed as well. A list of these stocks given in
terms of their stock symbol is presented in Table I along with the
starting date of the time-series and the total number of daily yields
in each. The ending date for all 24 time-series is December 29, 2006.    

\section{A Temporal Model of Stock Market Prices}
\label{sec: 2}

We begin our study of the temporal behavior of stock market prices
with an application to finance. Specifically, for the 24 stocks considered
here we study whether the daily yield on December 29, 2006 depends
on the daily yield on any day previous to it. This property of the market,
which has direct implications in finance, will be used as the starting
point for the construction of our model of stock prices.  

\subsection{An Inherent Contradiction}

\begin{figure}
\begin{center}
\includegraphics[width=1\textwidth, angle = 0]{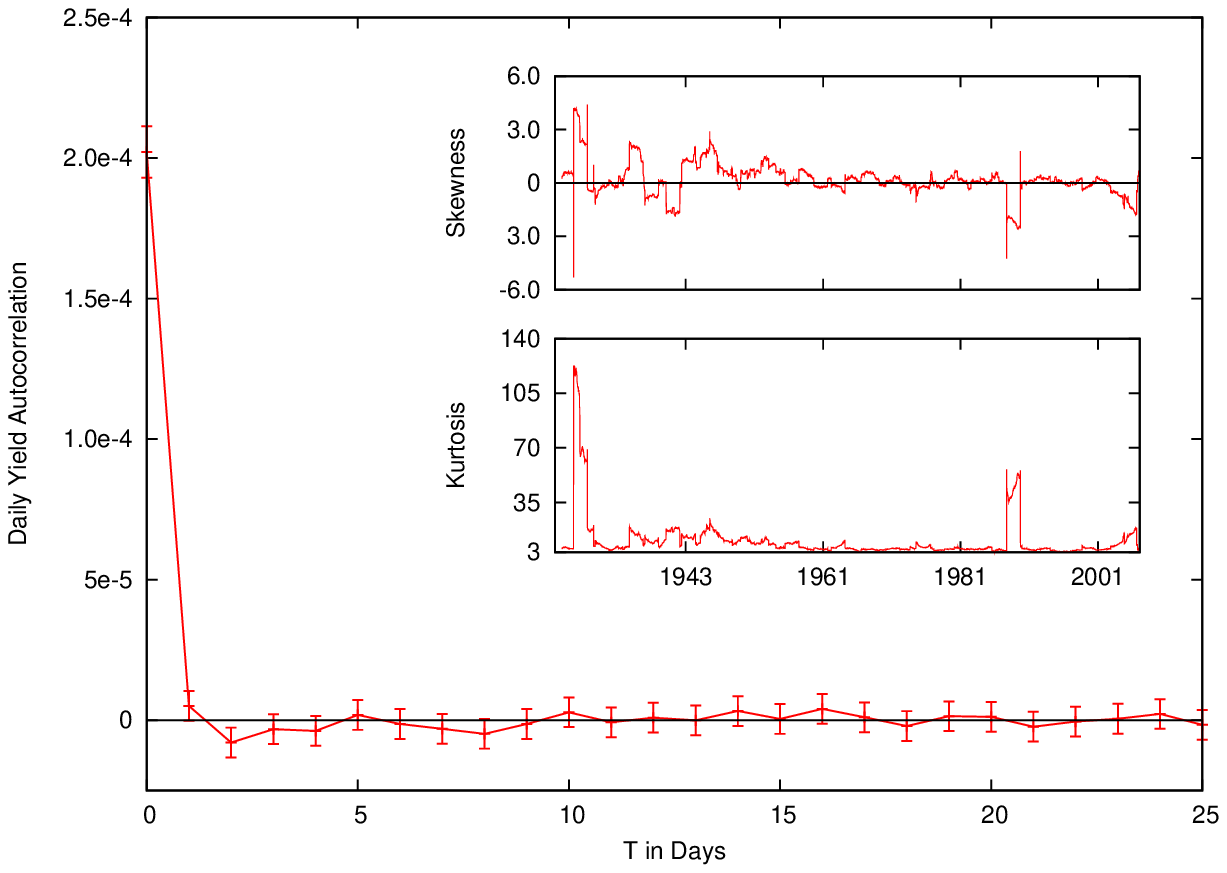}
Indications of an Efficiently Priced Stock
\end{center}
\caption{\label{Fig-1}
The autocorrelation of the daily yield for Coca Cola is shown in the main
figure, with the time, $T$, labeling the number of trading days
\textit{before} December 29, 2006. Also included at each data point
are errorbars set at $\pm$1.96 times the standard error. In the
insert, graphs of the sample skewness and the kurtosis of the stock
calculated using a 251-day moving average are shown. 
}
\end{figure}

Shown in Fig.~$\ref{Fig-1}$ is a graph of the autocorrelation
function of the daily yield for Coca Cola using
Eq.~$(\ref{GreenFunction})$ from the Appendix $\ref{sec: A2}$. This
autocorrelation is calculated between the daily yield of the stock on
December 29, 2006, and the daily yield $T$ days \textit{before} the
29th. The graph thus shows the dependency of the yield on the 29th on the
yield on any previous day.  If the yield on the 29th depends on
the yield on day $T$, then the autocorrelation function will not
vanish on that day at the 95\% CL. If, on the other hand, the yield on
the 29th does not depend on the yield on day $T$, then the
autocorrelation function will be within statistical error of zero. 

Also shown on the graph in Fig.~$\ref{Fig-1}$ is the errorbar for each
of the calculated values of the autocorrelation function.  These
errorbars are set at the 95\% CL, which is $1.96$ times the
standard error calculated using Eq.~$(\ref{varG})$ for the
autocorrelation function on that day. They thus set the 95\%
confidence interval (CI) about the calculated value for the
autocorrelation function.  If the value of the autocorrelation 
function falls within its errorbar of zero, there
is a 95\% probability that the autocorrelation on this day equals
zero. With 21,522 total trading days in the time-series for Coca
Cola, the standard error for the values of the autocorrelation function
shown in the graph is roughly 0.7\%, and is thus quite small; the
errorbars shown are correspondingly small. The standard error for the
majority of the stocks studied here are equally small.   

All but one of the errorbars for the autocorrelation shown in
Fig.~$\ref{Fig-1}$ straddles the $x$-axis. As such, we can say that the value
of the autocorrelation function for $T>0$ is within a 95\% CL of zero
for all but one day. Indeed, when we continue this calculation all the way
back to December 31, 1925, the starting date for the time-series, we 
find that the autocorrelation function for the daily yield on 20,279
out of a total of 21,522 trading days fall within the 95\% 
CI of zero; the autocorrelation function on 1,243 trading days, or 6\%
of the trading days, fall outside of the 95\% CI (see Table I). This
does not necessarily mean that there is a 
correlation between the 29th and these 1,243 trading days,
however. Statistically, we would expect values of the autocorrelation  
function to exceed the 95\% CI on 5\%, or 1,076, of the trading days. 
We can only conclude that on at least 1\%, or 215, of the trading days
the autocorrelation function does not vanish for $T>0$. If instead a
99\% CI in chosen, we find that the value of the autocorrelation
function falls within the 99\% CI of zero for 21,172 out of 21,522
trading days; they fall outside of the 99\% CI on only 2\%, or 350, of
the trading days. We can therefore still conclude that for at least
1\%, or 215, of the trading days the autocorrelation function may not
vanish for $T>0$.  

The autocorrelation function of the daily yield for all 24 stocks have
been calculated for the length their time-series, and we have found
that the autocorrelation function for these stocks behave 
similarly to Coca Cola's. Namely, the autocorrelation function is
maximum at $T=0$, and it does not vanish for at least 1\% to 3\% of
the trading days for each stock; for Citigroup and Verizon, this
percentage is even lower. We may conclude 
from this analysis that for the vast majority of the time the daily
yield of these stocks on any one day is not correlated with the daily
yield on any subsequent day; the market is thus extremely efficient
for these 24 stocks. In addition, we will show below that for the 1\% to
3\% of the trading days when the autocorrelation function does not
vanish, this is due to changes in the volatility of the stock with
time, and not to correlations between daily yields.  

Based on the above analysis, it would seem that the usual stochastic
process with a constant volatility would be a good model for these
stocks. The lack of dependence of the daily yield on any one day 
from any other is precisely the property inherent in such a
model. There are, however, other properties of the distribution of daily
yields for stocks that any model would have to explain as well, and it
is here that the constant-volatility model of stocks is lacking. 

Shown in the insert of Fig.~$\ref{Fig-1}$ is the sample skewness
of the daily yields for Coca Cola calculated using a
251-day moving average. If indeed the stock price of the stock is well
described by a stochastic process with a drift and a constant
volatility, then we would expect the skewness of the daily yield to be
zero. For Coca Cola, we find that the skewness ranges from $-5.3\pm
6.5$ to $4.4\pm 5.3$. Although the skewness is large in magnitude, its
standard error is correspondingly large, and we find that the skewness
exceeds the 95\% CI of zero on only 943 out of 21,022 days, or 4\%, of the
time. Thus, the sample skewness calculated using a 251-day moving
average agrees with what is expected from modeling the
yield of the stock using a stochastic process with  constant volatility.

The situation is quite different for the kurtosis, however. Shown also
in the insert of Fig.~$\ref{Fig-1}$ is the sample kurtosis
of the daily yields for Coca Cola calculated with the same
251-day moving average. Although the kurtosis for a daily yield
described by a stochastic process with a constant volatility is
expected to be three, what we find instead is that the sample 
kurtosis calculated for the Coca Cola time-series ranges in value from
$2.92\pm 0.22$ to $122\pm 49$.  Like the skewness, the standard error
for the kurtosis is large when the kurtosis is large, but unlike the
skewness, the error is not overwhelmingly large.  We find that the
kurtosis exceeded the 95\% CI of three on 15,393 out of 21,022 days, or
72\%, of the time. For the great majority of the trading days in the
time-series, the  kurtosis is different from that expected for a
stochastic process with constant volatility.   

We have done this calculation for all 24 stocks, and these results are
not unique to Coca Cola. This, then, is the contradiction inherent in using
a stochastic process with constant volatility to model stock market
prices. On the one hand, calculations of the autocorrelation function
indicate that the market is extremely efficient for these stocks, which
is consistent with a stochastic process with constant volatility. On
the other hand, calculation of the kurtosis for these stocks are much
larger than expected for such a process. We will use this
contradiction to guide the construction of our model in the analysis
below.

\begin{center}
\tablefirsthead{%
\hline\hline
 &$\quad\>$ Starting $\>\>$ &         & \multicolumn{2}{c}{Daily
  Yield}  & \multicolumn{2}{c}{Standardized Daily Yield}\\
\cline{4-7}
&$\quad\>$ Date $\quad\>$     &$\qquad N_T\>\,$ & $\quad >$ 95\% CI & $\quad >$ 99\% CI&  $\quad >$ 95\% CI &  $\quad >$ 99\% CI\\
\hline
}
\tablehead{%
\hline
\multicolumn{6}{l}{\small\sl continued from previous page}\\
\hline
 &$\quad\>$ Starting $\>\>$ &         & \multicolumn{2}{c}{Daily Yield}  & \multicolumn{2}{c}{Standardized Daily Yield}\\
\cline{4-7}
&$\quad\>$ Date $\quad\>$     &$\qquad N_T\>\,$ & $\quad >$ 95\% CI & $\quad >$ 99\% CI&  $\quad >$ 95\% CI &  $\quad >$ 99\% CI\\
\hline
}
\tabletail{%
}
\tablelasttail{\hline}
\topcaption{Autocorrelations for the 24 DJIA Stocks}     
\begin{supertabular}{lrrrrrr}
C     &10/29/86  &  5088  &   227 (4\%)  &   53 (1\%)  &  256 (5\%)   &   45 (1\%)  \\
MSFT  &03/13/86  &  5248  &   221 (4\%)  &   63 (1\%)  &  292 (6\%)   &   60 (1\%)  \\
VZ    &02/16/84  &  5770  &   236 (4\%)  &   58 (1\%)  &  306 (5\%)   &   57 (1\%)  \\
INTC  &12/14/72  &  8592  &   516 (6\%)  &  159 (2\%)  &  466 (5\%)   &   93 (1\%)  \\
AXP   &12/14/72  &  8592  &   412 (5\%)  &  101 (1\%)  &  391 (5\%)   &   78 (1\%)  \\
AIG   &12/14/72  &  8592  &   470 (5\%)  &  137 (2\%)  &  470 (5\%)   &   96 (1\%)  \\
{}    &          &        &              &             &              &             \\ 
WMT   &11/20/72  &  8608  &   458 (5\%)  &  118 (1\%)  &  415 (5\%)   &   85 (1\%)  \\
HPQ   &03/03/61  & 11524  &   610 (5\%)  &  169 (1\%)  &  575 (5\%)   &  112 (1\%)  \\
DIS   &11/12/57  & 12386  &   741 (6\%)  &  182 (1\%)  &  624 (5\%)   &  135 (1\%)  \\
AA    &06/11/55  & 14014  &   688 (5\%)  &  153 (1\%)  &  710 (5\%)   &  158 (1\%)  \\
MRK   &05/15/46  & 15444  &   946 (6\%)  &  233 (2\%)  &  835 (5\%)   &  190 (1\%)  \\
MMM   &01/15/46  & 15544  &   733 (5\%)  &  253 (2\%)  &  724 (5\%)   &  145 (1\%)  \\
{}    &          &        &              &             &              &             \\ 
JNJ   &09/25/44  & 15920  &   759 (5\%)  &  163 (1\%)  &  734 (5\%)   &  145 (1\%)  \\
PFE   &01/17/44  & 16126  &   812 (5\%)  &  181 (1\%)  &  812 (5\%)   &  181 (1\%)  \\
BA    &09/04/34  & 18946  &  1281 (7\%)  &  360 (2\%)  &  980 (5\%)   &  200 (1\%)  \\
CAT   &12/02/29  & 20358  &  1691 (8\%)  &  702 (3\%)  & 1027 (5\%)   &  239 (1\%)  \\
PG    &08/12/29  & 20442  &  1644 (8\%)  &  678 (3\%)  & 1026 (5\%)   &  223 (1\%)  \\
GE    &12/31/25  & 21518  &  1474 (7\%)  &  545 (3\%)  & 1104 (5\%)   &  242 (1\%)  \\
{}    &          &        &              &             &              &             \\ 
GM    &12/31/25  & 21518  &  1370 (6\%)  &  488 (2\%)  & 1120 (5\%)   &  251 (1\%)  \\
DD    &12/31/25  & 21520  &  1520 (7\%)  &  532 (2\%)  & 1071 (5\%)   &  215 (1\%)  \\
MO    &12/31/25  & 21520  &  1804 (8\%)  &  762 (4\%)  & 1047 (5\%)   &  242 (1\%)  \\
IBM   &12/31/25  & 21522  &  1387 (6\%)  &  461 (2\%)  & 1095 (5\%)   &  193 (1\%)  \\
KO    &12/31/25  & 21522  &  1243 (6\%)  &  350 (2\%)  & 1124 (5\%)   &  230 (1\%)  \\
XOM   &12/31/25  & 21526  &  1498 (7\%)  &  477 (2\%)  & 1096 (5\%)   &  220 (1\%)  \\
\end{supertabular}
\end{center}
\label{GreenSummary}

\subsection{The Continuous Model}

In this section, we show that for continuous stochastic models of stock
prices with a \textit{deterministic} volatility that changes with
time, the yield at time, $t$, does not depend on the 
yield at any other time, $t'$. Such a stock price is thus able
to model the properties of the autocorrelation function found for the 24
stocks above. In a later section, we will show that the
time-variation in the volatility can also explain the abnormally large
sample kurtosis.   

Take as the price of the stock at any time, $t$, the continuous
function $S(t)$. This is an approximation, of course. Stocks
are bought and sold in discrete time periods, and the prices
of these transactions are always recorded in discrete units. It is,
however, easier to develop an understanding of the model, and
to show a number of properties of it, using this continuous
approximation instead of using a discrete time-series of stock prices. In
the next section, when we develop a recursion relation for the
volatility, we will consider real-world data, and will discretize the
continuous model presented here. 

Our model for $S(t)$ is a stochastic process with a drift,
$\widetilde{\mu}(t)$, and a volatility, $\sigma(t)$, that change only
with time:    
\begin{equation}
\frac{1}{S}\frac{dS}{dt} = \widetilde{\mu}(t) + \sigma(t)\xi(t),
\label{S-eqn}
\end{equation}
where $\xi(t)$ is a Gaussian random variable such that 
\begin{equation}
E[\xi(t)] = 0, \qquad \hbox{and}\qquad E[\xi(t)\xi(t')] =\delta(t-t').
\label{delta}
\end{equation}
Here, $E[\xi]$ is the expectation value of $\xi$ over a Gaussian
distribution, and $\delta(t)$ is the Dirac delta function. We
emphasize that while Eq.~$(\ref{S-eqn})$ may have a form that is
similar to various stochastic volatility models of the stock market,
for us $\sigma(t)$ is a \textit{deterministic} function of time; it
does not have the random component that is inherent in stochastic
volatility models. 

As usual, it is more convenient to work with $u(t) \equiv
\ln[S(t)]$; for continuous compounding, $du/dt$ is then the
instantaneous yield of the stock. In terms of $u(t)$,
Eq.~$(\ref{S-eqn})$ reduces to
\begin{equation}
\frac{du}{dt} = \mu(t) + \sigma(t)\xi(t),
\label{stochastic}
\end{equation}
where $\mu(t) = \widetilde{\mu}(t) - \sigma^2(t)/2$. 

It is straightforward to show that for this stochastic process the
instantaneous yield at time, $t$, does not depend on the yield at any
other time, $t'$. To do so, consider the expectation value 
\begin{equation}
E\left[\left(\frac{du}{dt}\Bigg\vert_t-\mu(t)\right)
  \left(\frac{du}{dt}\Bigg\vert_{t'}-\mu(t')\right)\right] =
E[\sigma(t)\sigma(t')\xi(t)\xi(t')]. 
\end{equation}
Because $\sigma(t)$ is a deterministic function, it can be moved
outside the expectation value so that 
$E[\sigma(t)\sigma(t')\xi(t)\xi(t')] =
\sigma(t)\sigma(t')E[\xi(t)\xi(t')]$. Using Eq.~$(\ref{delta})$,
we then conclude that 
\begin{equation}
E\left[\left(\frac{du}{dt}\Bigg\vert_t-\mu(t)\right)
  \left(\frac{du}{dt}\Bigg\vert_{t'}-\mu(t')\right)\right] =
\sigma(t)^2\delta(t-t'),
\label{cont-auto} 
\end{equation}
so that the autocorrelation function of the instantaneous yield
vanishes unless $t=t'$; the yield of the stock at any one time does
not depend on the yield at any other time. Our model thus
describes a market for the stock that is efficient. This is 
to be expected. At each instant, $t$, Eq.~$(\ref{stochastic})$ describes a
Gaussian distribution with drift, $\mu(t)$ and volatility, $\sigma(t)$,
and it is well known that for a Gaussian distribution the daily yield
on any one day is not correlated with the daily yield on any other. 

Note that if the volatility was a function of $u$ as well as $t$, or
if it was itself a stochastic process, as it is taken to be in stochastic
volatility models, we could not have moved the volatilities outside
the expectation value to obtain Eq.~$(\ref{cont-auto})$. In these
cases, it is not clear whether the yield of the stock at any one
time depends on the yield at any other time.
 
Formally, the solution to Eq.~$(\ref{stochastic})$ is
straightforward. If $\sigma(t)>0$ for all $t$, divide through by 
$\sigma(t)$, and then reparametize time by taking
\begin{equation}
\tau = \int^t_0\sigma(s) ds.
\label{tau}
\end{equation}
Equation $(\ref{stochastic})$ then simplifies to
\begin{equation}
\frac{du}{d\tau} = \hat{\mu}(\tau) + \xi(\tau),
\label{normal}
\end{equation}
where 
\begin{equation}
\hat{\mu}(\tau(t)) - \mu(t)/\sigma(t)=0, 
\label{muRelation}
\end{equation}
and $\xi$ is still a Gaussian random variable, but now in
$\tau$. Equation $(\ref{normal})$ is simply a stochastic process with
drift $\hat{\mu}(t)$ and unit volatility; its solution in terms
of $\tau$ is well known. The solution to the original equation,
Eq.~$(\ref{stochastic})$, can then be obtained, at least in principle, by
integrating Eq.~$(\ref{tau})$, and then replacing $\tau$ with
resulting function of $t$.   

In practice, our task is much more difficult. We are \textit{not} 
given a $\sigma(t)$, and then asked to find the price, $S(t)$, of the
stock at subsequent times. We are instead given a collection 
of stock prices collected over some length of time, and
then asked to find the volatility. This is a much more difficult
problem, but surprisingly, it is a solvable one, as we will see in the
next subsection. 

\subsection{The Discrete Process and a Recursion Relation for $\sigma(t)$}

\begin{figure}
\begin{center}
\includegraphics[width=0.7\textwidth, angle = 270]{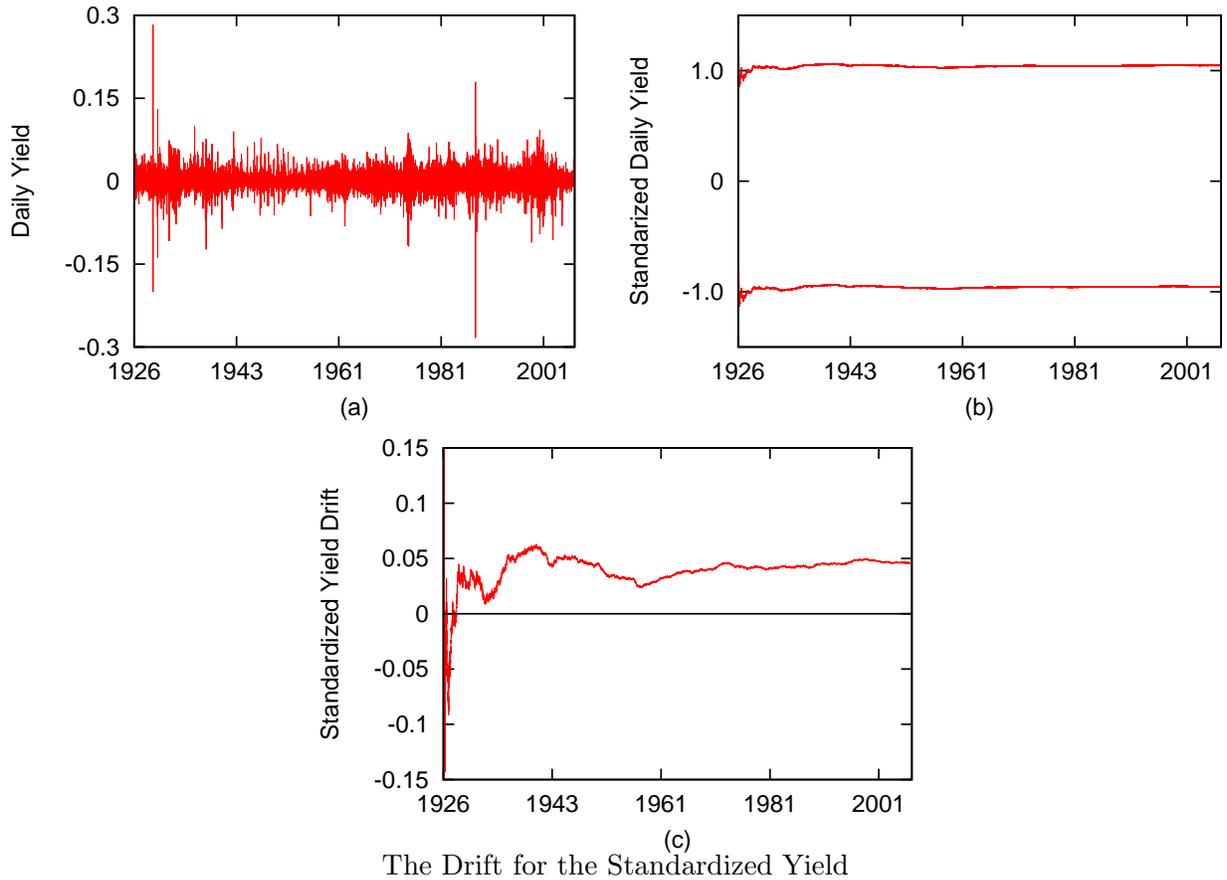}
The Drift for the Standardized Yield
\end{center}
\caption{\label{Fig-2}
A comparison between the distribution of daily and standardized daily
yields for Coca Cola is given in Figs.~2a and b. The binomial
behavior of the standardized daily yield can clearly be seen in
Fig.~2b. The resultant drift for the standardized daily yield is shown in
Fig. 2c.  
}
\end{figure}

In this subsection, we derive a recursions relation that is used to
solve for the volatility as a function of time. This derivation is
most conveniently done using a discretized version of the continuous
stochastic process Eq.~$(\ref{stochastic})$ considered above, and we
consider $S(t)$ as a continuous approximation to the discrete time
series, $S_n$, for $n = 1, \dots, N_T$, of stock prices collected at
equal time intervals, $a$; this $a$ is usually taken as one trading
day. The subscript $n$ enumerates the time step when the price of the
stock was collected, and is an integer that runs from 0 to the total
number of data points, $N_T$. As such, $t = n a$, $T=N_Ta$, 
$S_n\equiv S(na)$, $u_n \equiv u(na)$, and $\sigma_n \equiv \sigma(na)$ is
the volatility at $t=na$.  The instantaneous daily yield is then 
\begin{equation}
\frac{du}{dt} \approx \frac{u_n- u_{n-1}}{a} \equiv\frac{\Delta
  u_n}{a},
\end{equation}
where $n\ge 1$. It is clear that $\Delta u_n = \ln(S_n/S_{n-1})$ is
the yield of $S(t)$ over the time period $a$; when $a$ is one trading
day, $\Delta u_n$ is the daily yield. 

Our task in this section is to determine $\sigma_n$
\textit{given} the time-series $S_n$, and we do so by making use of the
analysis in the previous section. We call 
\begin{equation}
\frac{\Delta\hat{u}_n}{a} \equiv \frac{\Delta u_n}{a\sigma_n},
\label{SY}
\end{equation}
the \textit{standardized} yield of the stock price over a time period
$a$, and if $a$ is one trading day, we call it the standardized
daily yield. Since 
\begin{equation}
\frac{1}{\sigma(t)}\frac{du}{dt} \approx \frac{\Delta u_n}{a\sigma_n},
\label{defStand}
\end{equation}
then from the discretized versions of Eqs.~$(\ref{normal})$
and $(\ref{muRelation})$, we see that the distribution of standardized
yields has a volatility of $1/a$, or one, if $a$ is set to one
trading day. The collection of standardized yields has a \textit{known}
volatility. 

Consider now a subset of the time-series with $N<N_T$ elements, and
the corresponding collection of standardized yields,
$\Delta\hat{u}_n$, where $n$ runs now runs from $1$ to $N$. Because
this subset was arbitrarily selected from a collection of
standardized yields that has a volatility of $1/a$, this subset must
also have a volatility of $1/a$. As such 
\begin{equation}
\frac{1}{a}\left(1\pm\sqrt{\frac{2}{N}}\right) = \frac{1}{N-1}\sum_{n=1}^N
\left(\frac{\Delta u_n}{a\sigma_n}-\frac{1}{N}\sum_{m=1}^N \frac{\Delta
  u_m}{a\sigma_m}\right)^2,
\label{N}
\end{equation}
where we have included for
completeness in Eq.~$(\ref{N})$ the standard error for the volatility
given $N$ data points (see Stuart and Ord 1994) to
emphasize that the accuracy of Eq.~$(\ref{N})$ depends on
$N$. Equation $(\ref{N})$ must be true for each $N\le N_T$.  In
particular, it must hold for $N-1$, and thus we can write 
\begin{equation}
\frac{1}{a}\left(1\pm\sqrt{\frac{2}{N-1}}\right) =
\frac{1}{N-2}\sum_{n=1}^{N-1}
\left(\frac{\Delta u_n}{a\sigma_n}-\frac{1}{N}\sum_{m=1}^{N-1} \frac{\Delta
  u_m}{a\sigma_m}\right)^2,
\label{N-1}
\end{equation}
which is similar in form to Eq.~$(\ref{N})$. This self-similar
property of the distribution is used to determine $\sigma_n$, as we
show below.  

We first expand Eq.~$(\ref{N})$, and single out 
the $n=N$ terms,  
\begin{eqnarray}
\frac{1}{a} &=& \frac{1}{N-1}\left(\frac{\Delta u_N}{a\sigma_N}\right)^2
+ \frac{1}{N-1}\sum_{n=1}^{N-1}
\left(\frac{\Delta u_n}{a\sigma_n}\right)^2
\nonumber
\\
&{}& -
\frac{N}{N-1}\left(\frac{1}{N} \frac{\Delta
  u_N}{a\sigma_N}+\frac{N-1}{N^2}\sum_{m=1}^{N-1}\frac{\Delta
  u_m}{a\sigma_m}\right)^2,  
\label{N-2}
\end{eqnarray}
where we have dropped the error terms in Eq.~$(\ref{N})$ for clarity. Using
Eq.~$(\ref{N-1})$ in the second term of Eq.~$(\ref{N-2})$ and
completing a square, we arrive at a surprisingly simple equation for
$\sigma_N$,  
\begin{equation}
\left(\frac{N}{N-1}\right)\frac{1}{a} = \left(\frac{\Delta
  u_N}{a\sigma_N}-\frac{1}{N}\sum_{m=1}^{N-1}\frac{\Delta
  u_m}{a\sigma_m}\right)^2.
\label{recEq}
\end{equation} 
This is easily solved to give, 
\begin{equation}
\sigma_N \sqrt{a} = \Delta u_N \left(\frac{1}{N}\sum_{m=1}^{N-1}\frac{\Delta
  u_m}{a\sigma_m}\sqrt{a}\pm\sqrt{\frac{N}{N-1}}\right)^{-1},
\label{recursion}
\end{equation}
where the sign of the root must be chosen so that $\sigma_n > 0$ for all
$N$. The standardized yield, $\Delta\hat{u}_N$, can then be
calculated using Eq.~$(\ref{SY})$ for each time step. Equation
$(\ref{recursion})$ gives a recursion relation for $\sigma_N$. 

A recursive approach to calculating the volatility similar in
spirit to the one above is described in Stuart and Ord (1994). That
calculation is for volatilities that do not change with time, however,
while in ours the volatility can do so explicitly. As we will
see below, this introduces a number of complications. We note also that
Eq.~$(\ref{recursion})$ differs markedly from autoregression
approaches such as the EWMA, ARCH, and GARCH in that $\sigma_N$
depends nonlinearly on $\Delta u_N$.

Equation $(\ref{recursion})$ gives a first-order recursion relation for
$\sigma_n$, and thus given an initial $\sigma_1$, the values
for $\sigma_n$ for $n > 1$ is determined. To determine this initial
$\sigma_1$, we note that in the continuous process
Eq.~$(\ref{muRelation})$ holds. A similar relation must hold
for the discretized yields $\Delta u_n$. 

To determine this relation, we follow the same approach that led to
Eq.~$(\ref{N})$, and consider the following function 
\begin{equation}
f_N \equiv \frac{1}{N}\sum_{n=1}^N\frac{\Delta u_n}{a\sigma_n}-
\frac{1}{\sigma_N}\left(\frac{1}{N}\sum_{N=1}^{N_T}\frac{\Delta u_n}{a} \right).
\label{f_N}
\end{equation}
The first term in Eq.~$(\ref{f_N})$ is the average of the standardized
yield over the first $N$ terms in the time-series, and it corresponds to
the discretization of the first term in continuous constraint
Eq.~$(\ref{muRelation})$. The second term is the quotient of the
average daily yield calculated over the same period with the
volatility evaluated at the end of this period, and it corresponds to
the discretization of the second term in continuous constraint
Eq.~$(\ref{muRelation})$. If $\sigma_1$ can be chosen so that the mean
of $f_N$, 
\begin{equation}
\frac{1}{N}\sum_{n=1}^N f_N,
\label{avef}
\end{equation}
can be minimized to zero at the 95\% CL, then Eq.~$(\ref{muRelation})$
will hold on average for the discretized yield. As usual, the 95\% CL
for this mean is calculated through the standard error,
$[D(f_N)/N_T]^{1/2}$, where  
\begin{equation} 
D(f_N) \equiv\frac{1}{N_T-1}\sum_{N=1}^{N_T}\left(f_N -
\frac{1}{N_T}\sum_{M=1}^{N_T}f_M\right)^2,
\end{equation}
is the standard deviation of $f_N$.

We have successfully applied the recursion relation,
Eq.~$(\ref{recursion})$, to the 24 DJIA considered here, and have
obtained for each stock time-series for $\sigma_n$ and
$\Delta\hat{u}_n$. This was done by determining the quotient $\Delta
u_1/\sigma_1$ through an iterative search algorithm that was implemented
with a simple C++ program. This algorithm searches for a $\sigma_1$ that
drives the mean of $f_N$ to zero while in the process minimizing
$D(f_N)$. In addition, since the volatility must be non-negative, this
search is done under the constraint that all calculated values for
$\sigma_n$ must be greater or equal to zero, and it was stopped once
the mean of $f_N$ has been calculated to sufficient accuracy.   

A $\Delta u_1/\sigma_1$ that minimizes $f_N$
while at the same time giving a non-negative value for the volatility
can be found for all 24 stocks. Indeed, we found that the mean of
$f_N$ can be driven as close to zero as needed. The results of this 
calculation is given in Table II, which lists for each stock the value
of $\Delta u_1/\sigma_1$, the mean of $f_N$ for this $\Delta
u_1/\sigma_1$, and the standard error of the mean. While values for
$\Delta u_1/\sigma_1$ is only given to an accuracy of 
$10^{-7}$\textemdash which is sufficient given the accuracy of the
$S_n$ for the stocks as noted in Appendix $\ref{sec: A1}$\textemdash we
have been able to drive the mean value of $f_N$ to as far down as
$10^{-16}$ by increasing the accuracy of $\Delta u_1/\sigma_1$ to
$10^{-15}$. It is clear from the standard errors given in Table II
that the mean of $f_N$ vanishes within standard error at the 95\%
CL. This validates the recursion relation for $\sigma_n$ for all 24 stocks.   

Implicit in the derivation of Eq.~$(\ref{N})$ is that $N$ is
large, and yet since $\sigma_N$ starts at some initial point
$\sigma_1$, $\sigma_N$ are necessarily generated at small 
$N$. We would thus expect that there is a transient interval marked by
some $N_{tran}<N_T$ for which the solution to
Eq.~$(\ref{recursion})$ for $N< N_{tran}$ is markedly different
from the solution when $N> N_{tran}$. This is seen. For all 24 stocks,
the behavior of $\Delta\hat{u}_n$ for $n$ near one is different than
its behavior for large $n$.  This difference is similar for all of the
stocks, indicating that it is due to the recursion process itself, and not to
any underlying behavior of the markets. We would thus 
hesitate to use the calculated values for $\sigma_n$ when 
$n<N_{tran}$ to draw conclusions about the behavior of the stock. The
length of this interval, $N_{tran}$, varies from stock to 
stock, but typically ranges between 100 to 400 trading days. Given
that the shortest time-series considered here contains 5,088 trading
days, this interval is extremely short for all 24 stocks, and is not
relevant in practice. This is yet another reason why we have chosen
stocks that have a long track record to analyze. 

\begin{center}
\tablefirsthead{%
\hline\hline
     &  $\quad\quad\Delta u_1/\sigma_1\>\>\,$&  $\qquad$ Mean $f_N$   & $\quad$ SE for $f_N$  \\
\hline 
}
\tablehead{%
\multicolumn{4}{l}{\small\sl continued from previous page}\\
\hline
     &  $\quad\quad\Delta u_1/\sigma_1\>\>\,$&  $\qquad$ Mean $f_N$
  & $\quad$ SE for $f_N$  \\

\hline
}
\tabletail{
}
\tablelasttail{\hline}
\topcaption{Determining $\sigma_1$ for the 24 DJIA Stocks}   
\begin{supertabular}{lrrr}
GE   &  -1.29986151  &  3.70$\times 10^{-12}$  &  2.51$\times 10^{-6}$ \\
AXP  &  -1.11059569  & -4.56$\times 10^{-7}$  &  7.60$\times 10^{-6}$ \\
PFE  &  -0.69103252  &  6.37$\times 10^{-9}$  &  6.02$\times 10^{-6}$ \\
DIS  &  -0.62361094  & -1.66$\times 10^{-11}$  &  1.08$\times 10^{-5}$ \\
MSFT &  -0.23500530  & -4.61$\times 10^{-11}$  &  3.36$\times 10^{-5}$ \\
KO   &  -0.00633264  &  8.28$\times 10^{-12}$  &  3.39$\times 10^{-6}$ \\
{}   &               &                        &                      \\
PG   &   0.29288976  & -2.68$\times 10^{-5}$  &  2.47$\times 10^{-2}$ \\
GM   &   0.29925834  &  1.19$\times 10^{-11}$  &  5.08$\times 10^{-6}$ \\
AIG  &   0.30999620  & -2.73$\times 10^{-11}$  &  5.90$\times 10^{-6}$ \\
MMM  &   0.36915581  & -4.67$\times 10^{-11}$  &  5.32$\times 10^{-6}$ \\
AA   &   0.37469224  &  1.30$\times 10^{-12}$  &  4.77$\times 10^{-6}$ \\
HPQ  &   0.64209911  &  1.55$\times 10^{-11}$  &  6.24$\times 10^{-6}$ \\
{}   &               &                        &                      \\
JNJ  &   0.96565296  & -1.96$\times 10^{-11}$  &  5.54$\times 10^{-6}$ \\
CAT  &   0.98788045  & -2.20$\times 10^{-11}$  &  5.27$\times 10^{-6}$ \\
INTC &   0.99040135  & -6.99$\times 10^{-12}$  &  1.58$\times 10^{-5}$ \\
WMT  &   0.99093623  &  5.31$\times 10^{-8}$  &  1.53$\times 10^{-5}$ \\
C    &   1.05691370  &  1.41$\times 10^{-10}$  &  5.76$\times 10^{-6}$ \\
IBM  &   1.12634844  & -1.08$\times 10^{-11}$  &  5.73$\times 10^{-6}$ \\
{}   &               &                        &                      \\
MO   &   1.18101555  &  1.15$\times 10^{-11}$  &  1.35$\times 10^{-6}$ \\
VZ   &   1.30751627  &  2.03$\times 10^{-7}$  &  1.40$\times 10^{-5}$ \\
DD   &   1.34345528  & -4.88$\times 10^{-12}$  &  1.17$\times 10^{-5}$ \\
BA   &   1.41323601  & -2.74$\times 10^{-11}$  &  1.36$\times 10^{-5}$ \\
XOM  &   1.41414287  & -1.88$\times 10^{-7}$  &  5.08$\times 10^{-3}$ \\
MRK  &   1.41417225  & -9.05$\times 10^{-7}$  &  7.72$\times 10^{-3}$ \\
\hline
\end{supertabular}
\end{center}
\label{Recursion}

\section{The Standardized Drift and Its Distribution}
\label{sec: 3}

Although the recursion relation, Eq.~$(\ref{recursion})$, has been
successfully solved for all 24 stocks, we will delay
until Sec.~$\ref{sec: 4}$ to present the solutions to this equation.
Instead, we will first validate our model by showing that the stochastic
process introduced in the previous sections solves the overly large
kurtosis problem raised in Sec.~$\ref{sec: 2}$. In the process, we
will find that the distribution of standardized yield is a generalized
Rademacher distribution, and will show that the simple skewness and
kurtosis agree with the values for population skewness and kurtosis
for this distribution.  By doing so, we will also have validated our
model of stock market prices using the criteria outlined in
Sec.~$\ref{sec: 1}$. As part of this process, we will be able to
determine the drift of the yield as a function of time as well. 

\subsection{Observed Properties of the Standardized Yield}

To complement the autoregression calculation for the daily yield shown
in Fig.~$\ref{Fig-1}$, we
have calculated autocorrelation function for the standardized daily
yield, $G^{(2)}(\Delta\hat{u}_{N_T}, \Delta\hat{u}_{N_T-M})$, for all 24
DJIA stocks. A plot of $G^{(2)}(\Delta\hat{u}_{N_T}, \Delta\hat{u}_{N_T-M})$
as a function of $T=aM$ has the same shape as that shown in
Fig.~$\ref{Fig-1}$, but with $G^{(2)}(\Delta\hat{u}_{N_T},
\Delta\hat{u}_{N_T})=1$ for all the stocks instead of a range of
values. Like the daily yield, the standardized daily 
yield on any one day is not correlated with the standardized yield on
any other day; this is to be expected if the volatility is a function
of time only. We have also determined the number of trading days for
which the value of the autocorrelation function falls within the 95\%
and 99\% CI of zero. The results are shown in Table I, and for for all but
one stock, the results as as expected. On 5\% of the trading days, the value
of the autocorrelation function exceeds the 95\% CI, and on 1\% of the
trading days, the value exceeds the 99\% CI. The only exception is
Microsoft at the 95\% CL when on 6\% of the trading days the
autocorrelation function exceeds the 95\% CI of zero. 

In Sec.~$\ref{sec: 2}$, we noted that on at least 1\% of the trading
days the value of the autocorrelation function for the daily yield
exceeds either the 95\% or 99\% CI of zero, and we can say with a
degree of statistical certainty that for these days, the
autocorrelation function does not vanish. With the exception of
Microsoft at the 95\% CL, such days are not found
in the autocorrelation function of the standardized yields.
Since the standardized yield is obtained from the yield by removing
the time-dependent volatility, we conclude from the results in Table I
for the standardized yield that, with the possible exception of
Microsoft, this 1\% is not due to correlations in the daily yield, but
rather to temporal variations in the volatility.  

Next, shown in Fig.~$\ref{Fig-2}$a is a plot of the daily yield with
respect to trading day for Coca Cola. In comparison,
Fig.~$\ref{Fig-2}$b is the is the plot of the standardized daily yield
for the stock over the same period. It is readily apparent that
instead of taking a range of values between $\pm 0.3$ as the daily
yield does, the standardized yield jumps between two values, one near
$+1$ and one near $-1$. Notice also that while the standardized yield
is not precisely $+1$ or $-1$, any changes in the standardized yield 
near $+1$ are accompanied by the \textit{same} variations of
the yield near $-1$; the variations in the standardized yield near
$+1$ and near $-1$ would seem to move up or down in unison. Indeed, using a
251-day moving average, we find that the average of the difference in
the value of the standardized daily yield, $A_n^{(+)}$, near $+1$ and
its value, $A_n^{(-)}$,  near $-1$ ranges from a minimum of
$(A^{(+)}-A^{(-)})/2 = 0.9934\pm0.0061$ to a maximum of
$(A^{(+)}-A^{(-)})/2 = 1.0083\pm 0.0068$; both are within the 95\% CI
of one.  

This binomial behavior for the yield is not surprising for the same
reasons that binomial trees are effective at pricing options. As noted
by Cox and Ross (1976), a continuous stochastic process with constant
volatility can be approximated as a discrete random walk where at each
time step, $na$, there is a probability, $p$, that the stock price will
increase at the next time step, and a probability $1-p$ that it will
decrease. The discrete stochastic process can thus be approximated by
a binomial distribution, and as the binomial distribution is known to
approach the Gaussian distribution in the large $n$ limit, the
discrete random walk approaches a continuous stochastic process for
the stock price. Indeed, this limit is the reason why binomial
trees are effective in the first place.  

\subsection{Determining the Drift for the Standardized Yield}

In this subsection, we will determine the drift of the standardized
daily yield as a function of time. We do so by noting
that it is apparent from Fig.~$\ref{Fig-2}$ that the distribution of
standardized daily yields is a binomial distribution. Thus, at each
time step, $n$, there is a probability, $p$, that the standardized
yield will increase by an amount $A^{(+)}_n$ on that day, and
probability, $1-p$, that it will decrease by 
an amount $A^{(-)}_n$. While in principle $p$ may be different at
different time steps, the fact that $A^{(+)}_n$ and $A^{(-)}_n$ change
in unison while keeping the average distance between position and
negative standardized yields constant suggests that any variation in
time is due to an overall shift in the distribution. Variations in
$A^{(+)}_n$ and $A^{(-)}_n$ are not due to a time-dependent $p$,
but rather to a drift for the standardized yield that changes with
time. 

With this realization, the drift can easily be determined for all 24
stocks. From Eqs.$(\ref{normal})$ and $(\ref{defStand})$, we can
express the standardized daily yield as 
\begin{equation}
\Delta\hat{u}_n = \hat{\mu}_n +\xi^R_n,
\label{drift}
\end{equation}
where $\hat{\mu}_n\equiv \hat{\mu}(an)$ is the discretized drift of
the standardized yield, and $\xi^R_n$ is a random variable with zero
mean and unit volatility such that $E_R[\xi^R_n\xi^R_m] =
\delta_{nm}$. While for the continuous process $\xi_R$ would be a
Gaussian random variable, for the discrete process we will show that
$\xi^R_n$ is a random variable for the generalized Rademacher
distribution described below.   

The standard way of calculating $\hat{\mu}_n$ is to use a moving
average over a window of $M$ days. However, just like for the volatility,
calculating $\hat{\mu}_n$ with a moving average will mean that
variations in the drift faster than $M$ cannot be clearly
seen. We will instead calculate $\hat{\mu}_n$ directly from
$\Delta\hat{u}_n$, which is possible to do because the distribution
of standardized yield is so simple. 

We first note that since changes to $A^{(+)}_n$ and $A^{(-)}_n$ is due
to shifts in the distribution of standardized yield with time, these
shifts must be due to the drift, $\hat{\mu}_n$, of the standardized
yield. Shifts in random variables are trivial changes to the
distribution, however, and a drift that changes with time will not
materially change the distribution of standardized yields. 

We next note that $E_R[\xi^R_n] =0$. As the values of
$\Delta\hat{u}_n$ lie close to $\pm 1$, we conclude that $\xi^R_n$ can
only take the values $\pm1$. Any deviation by $A_n^{(\pm)}$ from $\pm
1$ must be due to the drift.  This drift can be determined by solving
the equation   
\begin{equation}
\hat{\mu}_n=\Delta\hat{u}_n - \xi^R_n,
\end{equation}
for $\hat{\mu}_n$ by taking the sign of $\xi^R_n$ to be the same
as the sign of $\Delta\hat{u}_n$. This solution is straightforwardly
implemented, and the results for Coca Cola is shown in
Fig.~$\ref{Fig-2}$c. For clarity, we have only shown the values of
the drift between $\pm 0.15$. While there are values that lie outside
of this range, they occur in the first 10 time steps in the series,
and are part of the transient behavior mentioned above. 

The drift for the standardized yield of all 24 stocks have be found
using this approach. Not surprisingly, we find that 21 out of the 24
stocks have a drift that positive for the great majority of the
time-series. What is surprising is that for three of the 24 stocks
(Exxon-Mobil, Merck, and Proctor and Gamble) the drift of the yield of
the stock is \textit{negative} outside the transient region. For these
stocks, the only reason why their price increases is due
to the random walk, and because the volatility is so
much greater than the drift.  

\subsection{The Distribution of Standardized Daily Yields}

While in the last subsection we determined the drift of the daily
yield, in this subsection we will show that the distribution of the
standardized daily yield is a generalized Rademacher
distribution shifted by the drift, $\hat{\mu}_n$. This will be done by
comparing the skewness and kurtosis of the Rademacher distribution
with the sample skewness and kurtosis for $\Delta\hat{u}_n$ after the
drift has been removed. We will show that the two agree at the
95\% CL, and doing so will both determine the distribution and validate
our model of stock market prices as a stochastic process with a
time-dependent volatility. We begin by describing the properties of
the generalized Rademacher distribution.  

A generalize Rademacher distribution consists of a random variable,
$\xi^R$, that takes the value $+1$ with probability $p$, and the value
$-1$ with probability $1-p$. We denote the expectation
value for this distribution as $E_{R}[\cdot]$, and find that the
population mean, $mom'_1=E_R[X]$, is simply 
\begin{equation}
mom'_1 = p-(1-p) = (2p-1).
\label{m1}
\end{equation}
This vanishes for $p = 1/2$. The $k$th population moment, $mom_k
\equiv E_R[(\xi^R-E_R[\xi^R])^k]$, is easily calculated to be 
\begin{equation}
mom_k = (2)^kp(1-p)\left[(1-p)^{k-1}+(-1)^kp^{k-1}\right].
\label{mk}
\end{equation}
The population variance is thus 
\begin{equation}
mom_2 =4 p(1-p)
\label{popVar}
\end{equation}
, while the
population skewness is 
\begin{equation}
Skew = \frac{1-2p}{\sqrt{p(1-p)}},
\label{skew}
\end{equation}
and the population kurtosis for the distribution is 
\begin{equation}
Kurt =\frac{1-3+3p^2}{p(1-p)}.
\label{kurt}
\end{equation}
Clearly, if $p = 1/2$, then $m_2 = 1$, $Skew = 0$, and
$Kurt = 1$; this is the Rademacher distribution, which is 
a special case of the binomial distribution. When $p \ne 1/2$, we call
this the generalized Rademacher distribution.

Given the plot in Fig.~$\ref{Fig-2}$b, we would expect that the distribution of
standardized yields to be a Rademacher distribution with $p=1/2$ at
all time steps. To show that that this is the case, we have calculated
the sample skewness and the kurtosis of the standardized yield after
the drift, $\hat{\mu}_n$, has been removed from
$\Delta\hat{u}_n$. This has been done for all 24 stocks using the
entire time-series for each. We then compared these sample skewness and
kurtosis with the population skewness and kurtosis for the Rademacher
distribution using the t-Test. For completeness, we have also calculated the
probability, $p$, for each stock by counting the total number of
$\Delta \hat{u}_n>0$, and compared it to the Rademacher value of
$p=1/2$ using the chi-squared test. The results of these calculations
and tests are given in Table III. We see that for all but four of
the stocks the fit is exceedingly good; the skewness, the kurtosis,
and the probability all agree at the 95\% CL. 

\pagebreak
\begin{center}
\tablefirsthead{%
\hline\hline
     & \multicolumn{2}{c} {Skewness}           & \multicolumn{2}{c} {Kurtosis}               & \multicolumn{2}{c} {Probability} \\
\cline{2-7}
     &  Mean$\quad\>\>\,$                       & t-Test $\>\>\>$ &  Mean$\quad\quad\>\>$               & t-Test $\>\>\>$ & $p\>\>\,$    &$\chi^2\>$\\
\hline 
}
\tablehead{%
\hline
\multicolumn{7}{l}{\small\sl continued from previous page}\\
\hline
     & \multicolumn{2}{c} {Skewness}           & \multicolumn{2}{c} {Kurtosis}               & \multicolumn{2}{c} {Probability} \\
\cline{2-7}
     &  Mean$\quad\>\>\,$                       & t-Test $\>\>\>$ &  Mean$\quad\quad\>\>$               & t-Test $\>\>\>$ & $p\>\>\,$    &$\chi^2\>$\\
\hline
}
\tabletail{%
\hline
\multicolumn{7}{l}{\small\sl continued on next page}\\
\hline
}
\tablelasttail{\hline}
\topcaption{
Model Validation
}    
\begin{supertabular}{lrrrrrr}
GM   $\quad$ &  $ 0.001 \pm 0.014$  &$\>\>\>$0.07$\>\>\>$  &  $1.00028 \pm 0.00030$  &$\>\>\>$0.92$\>\>\>$ &  0.500  &$\>\>\>$0.01\\
DD   $\quad$ &  $ 0.002 \pm 0.014$  &$\>\>\>$0.15$\>\>\>$  &  $1.00028 \pm 0.00031$  &$\>\>\>$0.92$\>\>\>$ &  0.499  &$\>\>\>$0.02\\
VZ   $\quad$ &  $-0.005 \pm 0.026$  &$\>\>\>$0.18$\>\>\>$  &  $1.0011\>\>  \pm 0.0012\>\> $  &$\>\>\>$0.92$\>\>\>$ &  0.501  &$\>\>\>$0.03\\
DIS  $\quad$ &  $ 0.010 \pm 0.018$  &$\>\>\>$0.54$\>\>\>$  &  $1.00058 \pm 0.00063$  &$\>\>\>$0.92$\>\>\>$ &  0.498  &$\>\>\>$0.29\\
AXP  $\quad$ &  $ 0.013 \pm 0.022$  &$\>\>\>$0.58$\>\>\>$  &  $1.00086 \pm 0.00093$  &$\>\>\>$0.92$\>\>\>$ &  0.497  &$\>\>\>$0.34\\
MMM  $\quad$ &  $-0.011 \pm 0.016$  &$\>\>\>$0.66$\>\>\>$  &  $1.00050\pm 0.00054$  &$\>\>\>$0.93$\>\>\>$ &  0.503  &$\>\>\>$0.43\\
{}           &                      &                      &                         &                     &         &        \\   
PG   $\quad$ &  $-0.009 \pm 0.014$  &$\>\>\>$0.67$\>\>\>$  &  $1.00038 \pm 0.00041$  &$\>\>\>$0.93$\>\>\>$ &  0.502  &$\>\>\>$0.45\\
JNJ  $\quad$ &  $-0.012 \pm 0.016$  &$\>\>\>$0.78$\>\>\>$  &  $1.00053 \pm 0.00056$  &$\>\>\>$0.94$\>\>\>$ &  0.503  &$\>\>\>$0.60\\
HPQ  $\quad$ &  $-0.015 \pm 0.019$  &$\>\>\>$0.78$\>\>\>$  &  $1.00072 \pm 0.00078$  &$\>\>\>$0.94$\>\>\>$ &  0.504  &$\>\>\>$0.61\\
KO   $\quad$ &  $-0.018 \pm 0.013$  &$\>\>\>$0.94$\>\>\>$  &  $1.00044 \pm 0.00046$  &$\>\>\>$0.96$\>\>\>$ &  0.503  &$\>\>\>$0.89\\
C    $\quad$ &  $-0.028 \pm 0.028$  &$\>\>\>$0.98$\>\>\>$  &  $1.0019\>\>  \pm 0.0020\>\> $  &$\>\>\>$0.97$\>\>\>$ &  0.507  &$\>\>\>$0.96\\
BA   $\quad$ &  $ 0.019 \pm 0.015$  &$\>\>\>$1.28$\>\>\>$  &  $1.00066 \pm 0.00064$  &$\>\>\>$1.04$\>\>\>$ &  0.495  &$\>\>\>$1.64\\
{}           &                      &                      &                         &                     &         &        \\   
AIG  $\quad$ &  $-0.028 \pm 0.022$  &$\>\>\>$1.32$\>\>\>$  &  $1.0015\>\>  \pm 0.0014\>\> $  &$\>\>\>$1.04$\>\>\>$ &  0.507  &$\>\>\>$1.73\\
INTC $\quad$ &  $-0.029 \pm 0.022$  &$\>\>\>$1.34$\>\>\>$  &  $1.0015\>\>  \pm 0.0015\>\> $  &$\>\>\>$1.05$\>\>\>$ &  0.507  &$\>\>\>$1.79\\
PFE  $\quad$ &  $-0.022 \pm 0.016$  &$\>\>\>$1.42$\>\>\>$  &  $1.00087 \pm 0.00081$  &$\>\>\>$1.07$\>\>\>$ &  0.506  &$\>\>\>$2.01\\
CAT  $\quad$ &  $-0.020 \pm 0.014$  &$\>\>\>$1.42$\>\>\>$  &  $1.00069 \pm 0.00064$  &$\>\>\>$1.07$\>\>\>$ &  0.505  &$\>\>\>$2.01\\
GE   $\quad$ &  $-0.020 \pm 0.014$  &$\>\>\>$1.49$\>\>\>$  &  $1.00069 \pm 0.00063$  &$\>\>\>$1.10$\>\>\>$ &  0.505  &$\>\>\>$2.21\\
AA   $\quad$ &  $ 0.025 \pm 0.170$  &$\>\>\>$1.50$\>\>\>$  &  $1.00107 \pm 0.00098$  &$\>\>\>$1.70$\>\>\>$ &  0.494  &$\>\>\>$2.26\\
{}           &                      &                      &                         &                     &         &        \\   
MSFT $\quad$ &  $-0.049 \pm 0.028$  &$\>\>\>$1.76$\>\>\>$  &  $1.0035\>\>  \pm 0.0030\>\> $  &$\>\>\>$1.19$\>\>\>$ &  0.512  &$\>\>\>$3.12\\
WMT  $\quad$ &  $-0.038 \pm 0.022$  &$\>\>\>$1.77$\>\>\>$  &  $1.0022\>\>  \pm 0.0018\>\> $  &$\>\>\>$1.19$\>\>\>$ &  0.510  &$\>\>\>$3.13\\
MRK  $\quad$ &  $-0.037 \pm 0.016$  &$\>\>\>$2.32$\>\>\>$  &  $1.0018\>\>  \pm 0.0013\>\> $  &$\>\>\>$1.40$\>\>\>$ &  0.509  &$\>\>\>$5.37\\
IBM  $\quad$ &  $-0.037 \pm 0.014$  &$\>\>\>$2.68$\>\>\>$  &  $1.0016\>\>  \pm 0.0010\>\> $  &$\>\>\>$1.55$\>\>\>$ &  0.509  &$\>\>\>$7.21\\
XOM  $\quad$ &  $-0.039 \pm 0.014$  &$\>\>\>$2.86$\>\>\>$  &  $1.0018\>\>  \pm 0.0011\>\> $  &$\>\>\>$1.63$\>\>\>$ &  0.510  &$\>\>\>$8.20\\
MO   $\quad$ &  $-0.040 \pm 0.014$  &$\>\>\>$2.94$\>\>\>$  &  $1.0019\>\>  \pm 0.0011\>\> $  &$\>\>\>$1.66$\>\>\>$ &  0.510  &$\>\>\>$8.67\\
\hline
\end{supertabular}
\end{center}
\label{Statistics}
\vskip 12pt

Although this agreement is not as good for Altria, Exxon, IBM, and
Merck, this is only because we were comparing it with the 
Rademacher distribution with $p=1/2$. We find that the distribution of
standardized yields for these stocks is instead the generalized Rademacher
distribution with a $p$ slightly greater than $1/2$. Using
Eq.~$(\ref{m1})$ and the sample mean for these stocks, we have solved
for a probability, $p$, for the stocks. We find that this probability
is in close agreement with those listed in Table III for these stocks,
and when this $p$ is then used in Eqs.~$(\ref{skew})$ and
$(\ref{kurt})$ to predict values for the skewness and kurtosis, the
predicted values are now in agreement with the sample skewness and
kurtosis at the 95\% CL. 

We also note that variance of the distribution calculated using the
values for $p$ given in Table 4 in Eq.~$(\ref{popVar})$ ranges from
0.9994 to 1.000. This is in excellent agreement with the requirement
that the variance of the standardized daily yield is one when $a$ is
one trading day. 

We thus conclude that the distribution of standardized daily yields
is a generalized Rademacher distribution shifted by the drift,
$\hat{\mu}_n$. For 20 of these stocks, we find that $p =1/2$ at the
95\% CL. The probability that the daily yield increases is the same
as the probability that it decreases. For the other four stocks, $p$
is slightly greater than $1/2$, and the probability that the daily
yield increases is slightly larger than the probability that it
decreases.
  
\section{The Instantaneous Volatility}
\label{sec: 4}

Having determined the distribution for the standardized daily yield,
we now turn our attention to determining the volatility of the
stock. 

\begin{figure}
\begin{center}
\includegraphics[width=0.71\textwidth, angle = 270]{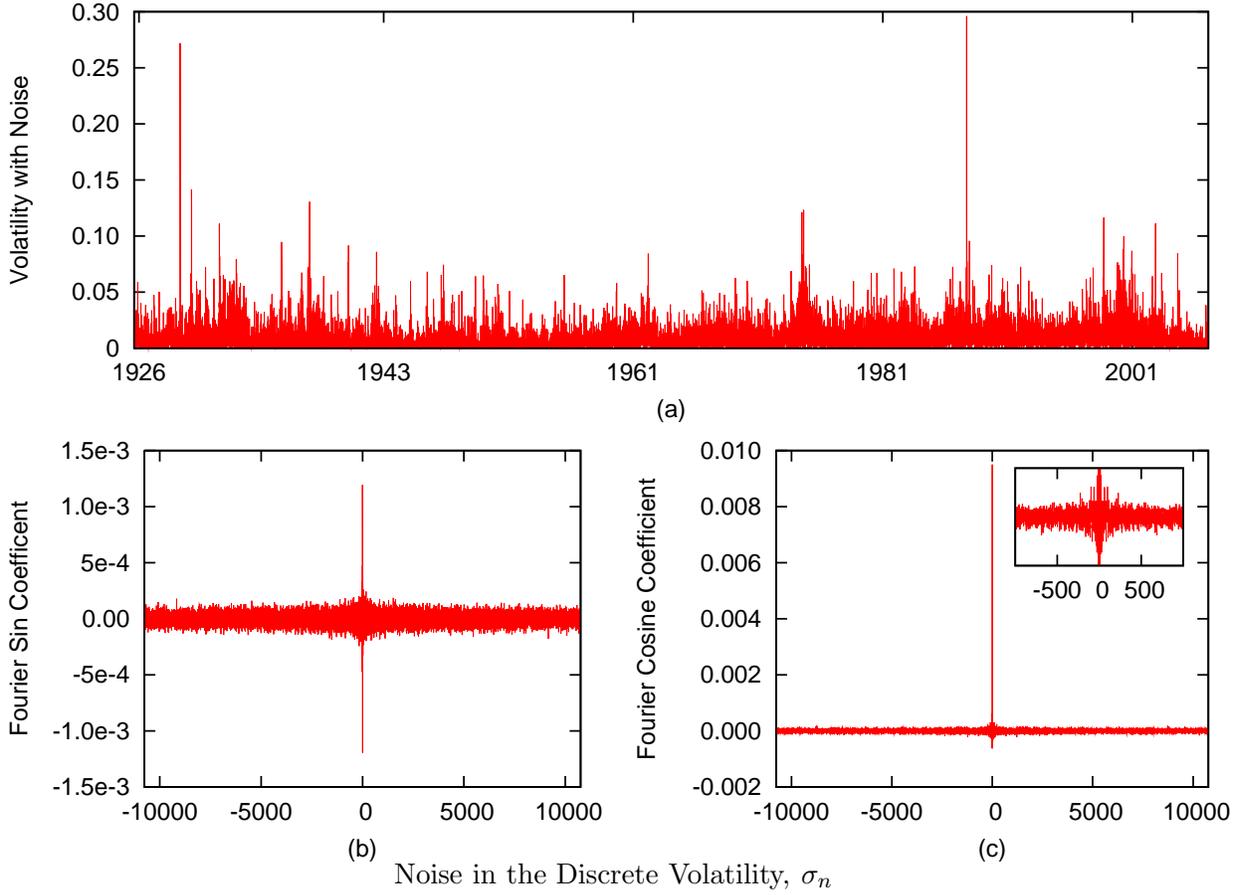}
Noise in the Discrete Volatility, $\sigma_n$
\end{center}
\caption{\label{Fig-5}
The top figure shows the volatility for Coca Cola obtained from
the recursion relation Eq.~$(\ref{recursion})$. The high degree
of noise associated with this volatility can readily be seen. In 
Figs.~$\ref{Fig-5}$b and c, the Fourier sine and cosine components
are graphed, and the floor of noise for both can readily be seen along
with the points that are above the noise.  From the insert in Fig.~3c,
the similarity in the noise floors for the sine and cosine coefficients 
is apparent.}  
\end{figure}

We find that while the recursion relation, Eq.~$(\ref{recursion})$, is
straightforwardly solved using the $\sigma_1$ given in Table II, there
is a great deal of noise associated with the resultant values for
$\sigma_n$. This can be seen in Fig.~$\ref{Fig-5}$a where we have
plotted as a function of trading day the volatility obtained from
Eq.~$(\ref{recursion})$. Although we can discern that there is an 
inherent structure in graph, this structure is buried within random
fluctuations of $\sigma_n$. These fluctuations are due to 
random noise generated when Eq.~$(\ref{recursion})$ is solved,
and they mask the functional dependence of $\sigma$ on $t$. In this
section, we will extract this dependence from the noise. 

The presence of the noise in $\sigma_n$ is inherent, but not because
$\sigma(t)$ itself obeys a stochastic process, as is assumed in stochastic
volatility models. If it were, then there will necessarily be a second
stochastic differential equation for $\sigma(t)$ to augment
Eq.~$(\ref{normal})$, and the two coupled equations would have to be
solved simultaneously. Certainly, Eq.~$(\ref{normal})$ and the recursion
relation Eq.~$(\ref{recursion})$ would not, in general, be solutions
of the coupled stochastic differential equations, and it is this
recursion relation that was used to obtain
Fig.~$\ref{Fig-5}$a. Rather, this noise is inherent in determining the
volatility itself. 

Note from Eq.~$(\ref{recursion})$ that $\sigma_n\propto \Delta
u_n/a$. For a stochastic process of the form Eq.~$(\ref{stochastic})$
where the volatility changes with time, at each time step, $an$,
$\Delta u_n$ is a random variable from a distribution with volatility 
$\sigma_n$. As $\sigma_n$ need not equal $\sigma_m$ for any two $n$
and $m$, each $\Delta u_n$ can come from a \textit{different}
distribution. In the worst case, we will have only \textit{one}
$\Delta u_n$ out of any distribution with which to determine
$\sigma_n$, and this $\Delta u_n$ can take any value from $-\infty$
to $+\infty$ with a probability 
\begin{equation}
P(\Delta u_n/a) = \frac{1}{\sigma_n}\sqrt{\frac{a}{2\pi}}e^{-(\Delta
  u_n/a-\mu_N)^2a/2\sigma_n^2}. 
\end{equation}
 Determining $\sigma_n$ would thus seem to be an impossible task. That it
 can nevertheless be done is due to three observations. First, because
 $P(\Delta u_n/a)$ is Gaussian, there is a 68\% probability that any
 value of $\Delta u_n/a$ will be within $\mu_n \pm
 \sigma_n/\sqrt{a}$. It is for this reason that it is still possible
 to discern an overall functional dependence of $\sigma_n$ on the
 trading day through the noise in
 Fig.~$\ref{Fig-5}$. Second, $\sigma(t)$ is a \textit{deterministic}
 function of $t$, and thus the value of the volatility at time step
 $an$ is related to its value at time step $a(n-1)$. Given a
 sufficient number of $\Delta u_n$\textemdash and thus a sufficient
 number of $\sigma_n$\textemdash it must be possible to construct a
 functional form for $\sigma(t)$. Third,  using Fourier analysis (also
 called spectral analysis) and signal  processing techniques, it is
 possible to remove from Fig.~$\ref{Fig-5}$ the noise that is
 obscuring the details of how  $\sigma_n$ depends on the trading day,
 and obtain a functional form for the volatility.  

That Fourier analysis provides an efficient way of removing the noise from
Fig.~$\ref{Fig-5}$ is based on the following theorem:

\textbf{Theorem:} If $\{\xi_n : n = 1, \dots, N\}$ is a time
series where $\xi_n$ is a Gaussian random variable with zero mean, and 
volatility, $\sigma$, then the Fourier sine,
$\alpha^{\sin}_k$,  and Fourier cosine, $\alpha^{\cos}_k$, coefficients
of the Fourier transform of $\xi_n$ are Gaussian random variables with
zero mean and volatility $\sigma/\sqrt{N}$. 

This theorem is well-known in signal analysis, and is an immediate
consequence of Parseval's Theorem. A proof of this theorem, as well as
a review of the discrete Fourier transform, is given in Appendix
$\ref{sec: A3}$. It is because the volatility of the Fourier
sine and cosine coefficients for Gaussian random variables are reduced
by a factor of $1/\sqrt{N}$ that it is possible to remove from $\sigma_n$
the random noise. In general, this reduction in the coefficients does
not occur if the $\xi_n$ are \textit{not} random variables, and thus
the structure in Fig.~$\ref{Fig-5}$ can be resolved once the Fourier
transform of $\sigma_n$ is taken. After this removal is accomplished, we
can then take the inverse Fourier transform to obtain $\sigma(t)$,
which we call the instantaneous volatility to differentiate it from
the $\sigma_n$ that comes directly from Eq.~$(\ref{recursion})$. 

Figure $\ref{Fig-5}$b and c are plots of the Fourier sine and Fourier
cosine coefficients of the discrete Fourier transform of $\sigma_n$
defined as 
\begin{equation}
a^{\cos}_k = \frac{1}{N_T}\sum_n^{N_T} \sigma_n \cos\left(\frac{2\pi
  ikn}{N_T}\right), \qquad 
a^{\cos}_k = \frac{1}{N_T}\sum_n^{N_T} \sigma_n \sin\left(\frac{2\pi
  ikn}{N_T}\right).
\end{equation} 
They depend on an integer $k$, which runs from
$-(N_T-1)/2$ to $(N_T-1)/2$. As 
\begin{equation}
\sigma_n = a_0^\omega +
2\sum_{k=1}^{(N_T-1)/2}a_k^{\cos}\cos\left(\frac{2\pi
  kn}{N_T}\right)+
2\sum_{k=1}^{(N_T-1)/2}a_k^{\sin}\sin\left(\frac{2\pi
  kn}{N_T}\right),
\label{signal}
\end{equation}
the Fourier transform decomposes the time-series, $\sigma_n$, into
components that oscillate with frequency $f_k = k/N_T$
day$^{-1}$ (or, equivalently, with period $N_T/k$ days) for
$k>0$; the coefficients $2\vert a^{\cos}_k\vert$ and $2\vert
a^{\sin}_k\vert$ are the amplitudes of these oscillations. 

In the graphs shown in Figs.~$\ref{Fig-5}$b and$\ref{Fig-5}$c, we can
readily see that there is a component of the Fourier coefficients for
Coca Cola that varies randomly between $\pm 0.0002$. This is the noise
floor. Coefficients in this floor are the result of the Fourier
transform of the noise that mask the functional behavior of $\sigma_n$
on the trading. This noise floor is similar for both the Fourier sine and
cosine coefficients, as can be see in detail in inset plot in
Fig.~$\ref{Fig-5}$c where the features of the plot of $a^{\cos}_k$ are
magnified for $k$ between $\pm 1000$.

It is also apparent from the graph that there are Fourier coefficients that
rise above the noise. While the most prominent of these is
$a_0^{\cos}$ (which is the average of $x_n$ over all trading days),
such points exist for other coefficients as well. This is due 
to the structure in $\sigma_n$ shown in Fig.~$\ref{Fig-5}$a; if there
were no structure at all in the plot, then there would be no Fourier
coefficients that rise above the noise floor. 

\begin{figure}
\begin{center}
\includegraphics[width=1\textwidth, angle = 0]{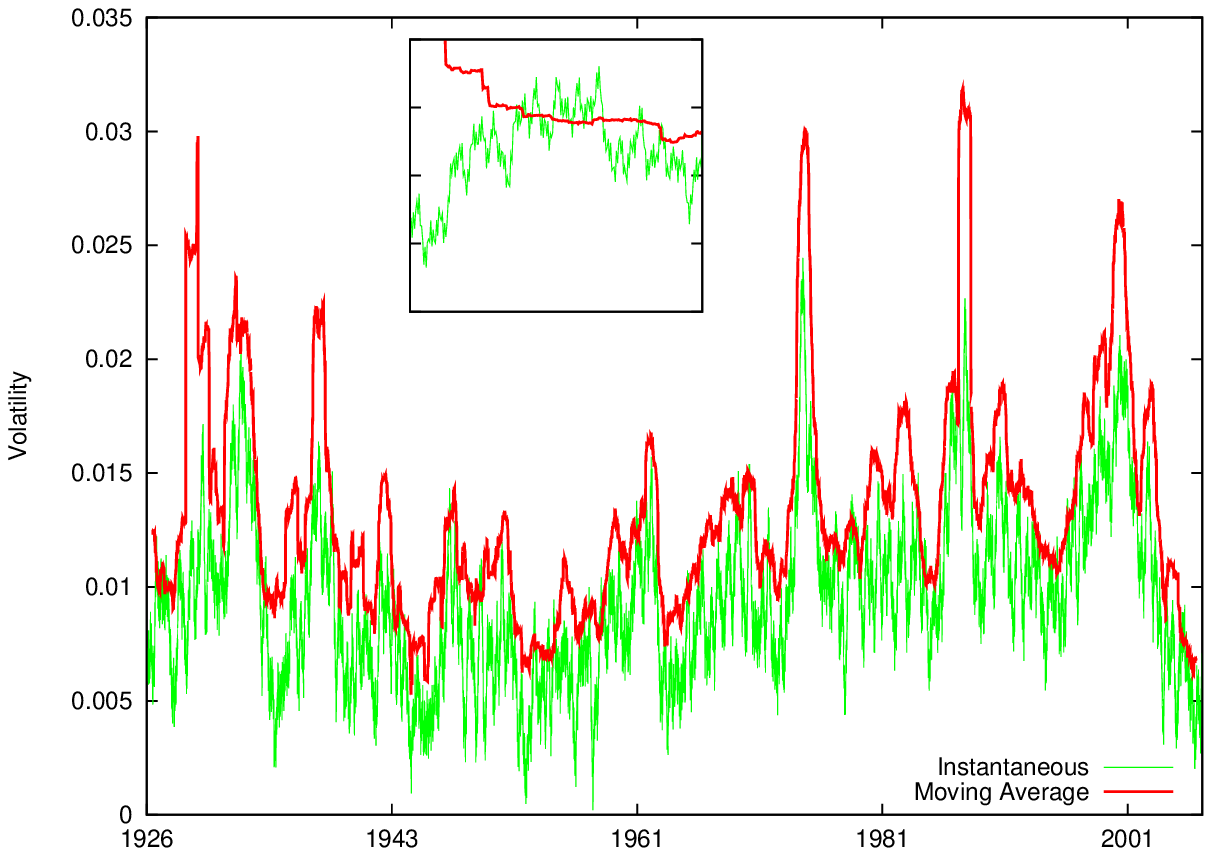}
The Instantaneous Volatility, $\sigma(t)$, After Noise Removal 
\end{center}
\caption{\label{Fig-6}
In the main figure, the instantaneous volatility Coca Cola and the
historical volatility the stock calculated with a 251-day moving
average is plotted. The historical volatility consistently
overestimates the instantaneous volatility. The degree of this
overestimation, along with the details that the moving average misses,
can be readily seen in the inset graph where both volatilities are plotted
over a one-year span from December 2, 2004 to December 01, 2005. 
}
\end{figure}

By combining this observation with the near uniformity of
the noise floor, we  are able to filter out the noise component of
$\sigma_n$, and construct a (approximately) noise-free instantaneous
volatility, $\sigma(t)$. A description of the process that we used,
along with the statistical criteria used to determine the noise floor
for the Fourier sine and cosine coefficients, is given in detail in Appendix 
$\ref{sec: A3-noise}$. The effectiveness of the noise removal process
can be seen in Fig.~$\ref{Fig-6}$ where a plot of the instantaneous
volatility for  Coca Cola is shown. When this plot is compared to
Fig.~$\ref{Fig-5}$a, the amount of noise removed, and the
success of the noise removal procedure, is readily
apparent. Indeed, out of a total of 21,523 Fourier sine and cosine
coefficients for $\sigma_n$, 10,734 Fourier sine and 10,728 of Fourier
cosine coefficients were removed as noise; only 59 points were kept to
construct $\sigma(t)$. While the graph of $\sigma(t)$ may appear
to be noisy, this is because eight decades of trading days are 
plotted in the figure. Much of this apparent noise disappears when the
range of trading days plotted is narrowed, as can be seen in the
inset figure. Here, the instantaneous volatility over a one-year
period from December 29, 2005 to December 29, 2006 has been plotted. 

To compare the instantaneous volatility with the historical
volatility, we have included in Fig.~$\ref{Fig-6}$ a graph of
historical volatility calculated from the daily yield using a 251-day
moving average. It is immediately apparent that the historical
volatility is generally larger than the instantaneous volatility; at
times it is dramatically so. It is also readily apparent that the historical
volatility does not show nearly as much detail as the instantaneous
volatility, as can be seen in the inset figure. 

A functional form for $\sigma(t)$ can be found for all 24 stocks. For
Coca Cola, this expression has 59 terms; we give only four of them here,
\begin{eqnarray}
\sigma(t) &=& 0.00950\Bigg[ 1  -0.25126\sin(2\pi f_0 t) +
  0.12670\cos(2\pi f_0 t) 
\nonumber
\\
&{}&\qquad\qquad + , \dots, +\> \hbox{55 terms}\> +, \dots, +\>
0.03870\cos(8735[2\pi f_0] t) \Bigg],
\label{instantaneousSigma}
\end{eqnarray}
where $f_0 = 1/21522$ rad/day is the fundamental angular
frequency. The amplitude of the first term in the expression is the
largest; it is the average of $\sigma_n$ over all the trading days in
the time-series. The second largest amplitude is the sine
term in Eq.~$(\ref{instantaneousSigma})$, and it is 25\% the size of
the  first. All other amplitudes are smaller then this term, for most
by a factor of 5, and yet notice from Fig.~$\ref{Fig-6}$ that these
amplitudes are nonetheless sufficient to generate a instantaneous 
volatility that is far from a constant function. 

From the last term in Eq.~$(\ref{instantaneousSigma})$, we see the
that shortest frequency of oscillations that make up $\sigma(t)$ is
$8735/21522\approx 0.4$ day$^{-1}$. This is very close to the Nyquist
criteria of $0.5$ day$^{-1}$ for $\sigma(t)$, which is the upper limit
on the frequencies of the Fourier components of $\sigma(t)$. The
underlying reason for such a limit is because the the
original time-series, $S_n$, was acquired once each trading day. We
therefore cannot \textit{measure} oscillations with a period shorter 
than two trading days; there simply is not enough information about
the stocks to determine what happens within the trading
day. (This is in contrast to \textit{predicting} how the volatility
may behave during the trading day, which certainly can be done.) For
each of the 24 stocks, the shortest period of the Fourier 
components that make up the instantaneous volatility are listed in
Table III, and we see that for all but 3 of the stocks our expression
for $\sigma(t)$ comes very close to Nyquist criteria. In the case
of Alcoa, Caterpillar, and Johnson \& Johnson, the shortest period has
even reached it.   
\pagebreak

\begin{center}
\tablefirsthead{%
\hline\hline
     &   & \multicolumn{2}{c}{$a_k^{\sin}$ Noise} & \multicolumn{3}{c}{$a_k^{\cos}$ Noise}\\
\cline{3-7}
     &$\qquad$ Period &$\qquad\quad$  Floor  & $\quad$ Kurtosis & $\qquad\quad$ Floor  &$\quad$  Skewness &$\quad$ Kurtosis  \\
     &$\qquad$ (days) &  $\qquad\quad\times$ SD  &  &  $\qquad\quad \times$ SD & &  \\
\hline 
}
\tablehead{%
\hline
\multicolumn{7}{l}{\small\sl continued from previous page}\\
\hline
     &   &\multicolumn{2}{c}{$a_k^{\sin}$ Noise} & \multicolumn{3}{c}{$a_k^{\cos}$ Noise}\\
\cline{3-7}
     &$\qquad$ Period &$\qquad\quad$  Floor  & $\quad$ Kurtosis & $\qquad\quad$ Floor  &$\quad$  Skewness &$\quad$ Kurtosis  \\
     &$\qquad$ (days) &  $\qquad\quad\times$ SD  &  &  $\qquad\quad \times$ SD & &  \\
\hline
}
\tabletail{%
\hline
\multicolumn{7}{l}{\small\sl continued on next page}\\
\hline
}
\tablelasttail{\hline}
\topcaption{
Effectiveness of Noise Filtering Routine
}     
\begin{supertabular}{lrrrrrr}
CAT  &  2.0&  3.33&  3.07&  3.36&  -0.04&  3.04\\
JNJ  &  2.0&  3.20&  2.99&  3.23&  -0.03&  2.99\\
AA   &  2.0&  3.40&  3.00&  3.47&  -0.02&  3.00\\
GE   &  2.1&  3.47&  3.01&  3.44&  -0.04&  3.00\\
HPQ  &  2.1&  3.45&  3.00&  3.31&  -0.08&  3.01\\
DIS  &  2.1&  3.42&  2.99&  3.64&  -0.03&  3.03\\
{}   &     &      &      &      &       &      \\
XOM  &  2.2&  3.79&  3.11&  3.61&  -0.05&  3.03\\
MSFT &  2.3&  3.51&  2.99&  3.27&  -0.08&  3.00\\
PFE  &  2.3&  3.38&  3.00&  3.40&  -0.05&  3.00\\
AXP  &  2.4&  3.56&  3.17&  3.33&  -0.04&  3.14\\
MRK  &  2.4&  3.33&  3.03&  3.95&  -0.03&  3.00\\
INTC &  2.4&  3.26&  3.00&  3.36&  -0.05&  2.99\\
{}   &     &      &      &      &       &      \\
WMT  &  2.5&  3.47&  2.96&  3.29&   0.11&  2.99\\
KO   &  2.5&  3.89&  3.30&  3.53&   0.02&  3.23\\
BA   &  2.7&  3.37&  3.00&  3.30&  -0.01&  2.99\\
PG   &  3.2&  3.31&  2.95&  3.40&  -0.04&  3.00\\
AIG  &  3.2&  3.44&  3.00&  3.32&   0.01&  3.00\\
MMM  &  3.5&  3.03&  2.54&  3.06&  -0.04&  2.58\\
{}   &     &      &      &      &       &      \\
IBM  &  4.0&  3.41&  3.00&  3.64&   0.03&  3.00\\
VZ   &  5.2&  3.28&  2.99&  3.23&  -0.09&  3.00\\
MO   &  5.5&  3.56&  3.14&  3.64&  -0.07&  3.21\\
C    &  6.2&  3.36&  3.02&  3.33&   0.05&  3.12\\
DD   &  9.6&  3.81&  3.31&  3.64&   0.00&  3.12\\
GM   & 14.2&  3.59&  3.00&  3.64&  -0.06&  3.00\\
\hline
\end{supertabular}
\end{center}
\label{FourierSummary}
\vskip 12pt

\begin{figure}
\begin{center}
\includegraphics[width=0.71\textwidth, angle = 270]{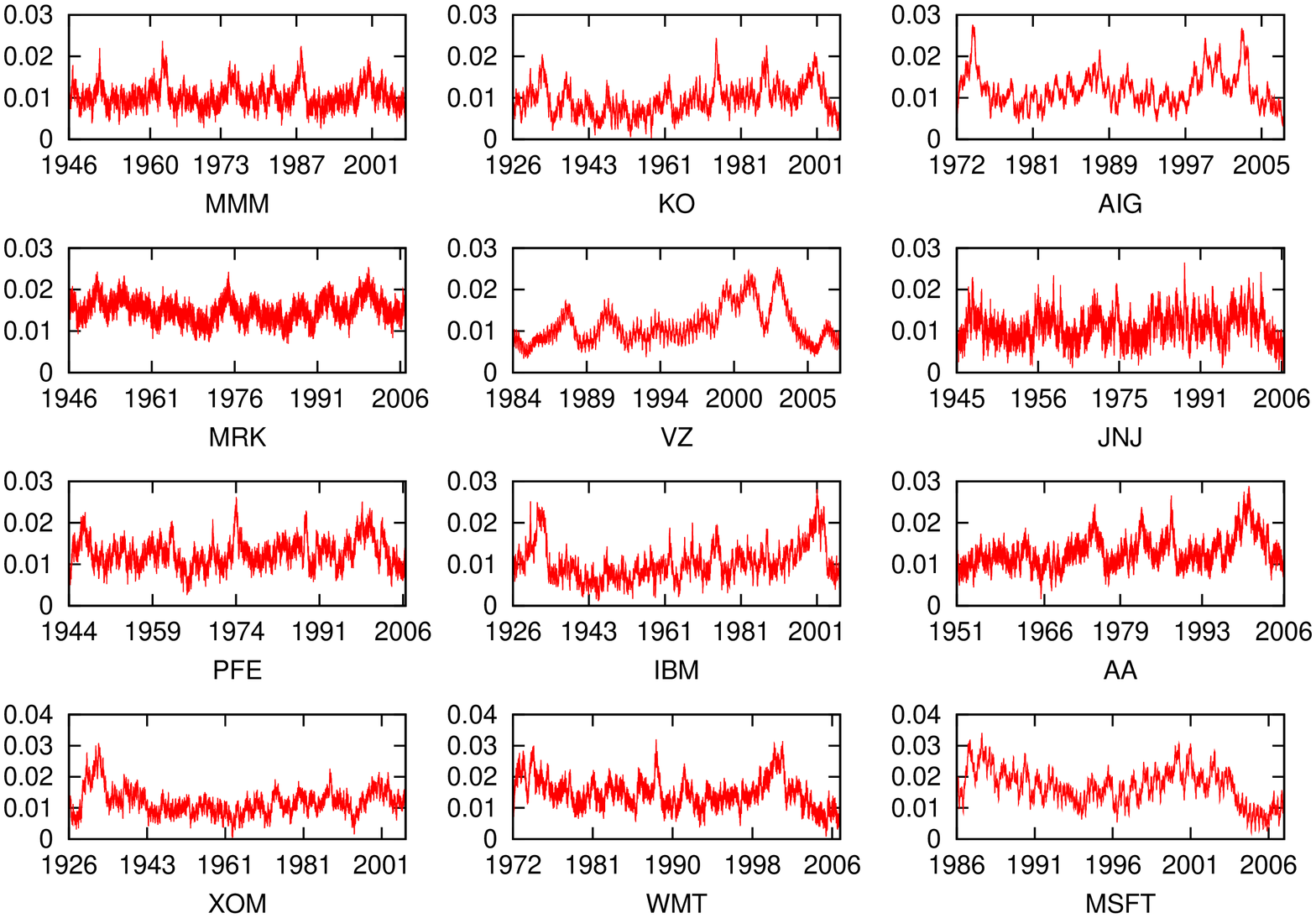}
The Instantaneous Volatility for the DJIA Stocks, I
\end{center}
\caption{\label{Fig-7}
Graphs of the instantaneous volatilities verses trading day for the 12 of
the 24 DJIA stocks with the lowest peak volatility are shown.
}
\end{figure}

\begin{figure}
\begin{center}
\includegraphics[width=0.71\textwidth, angle = 270]{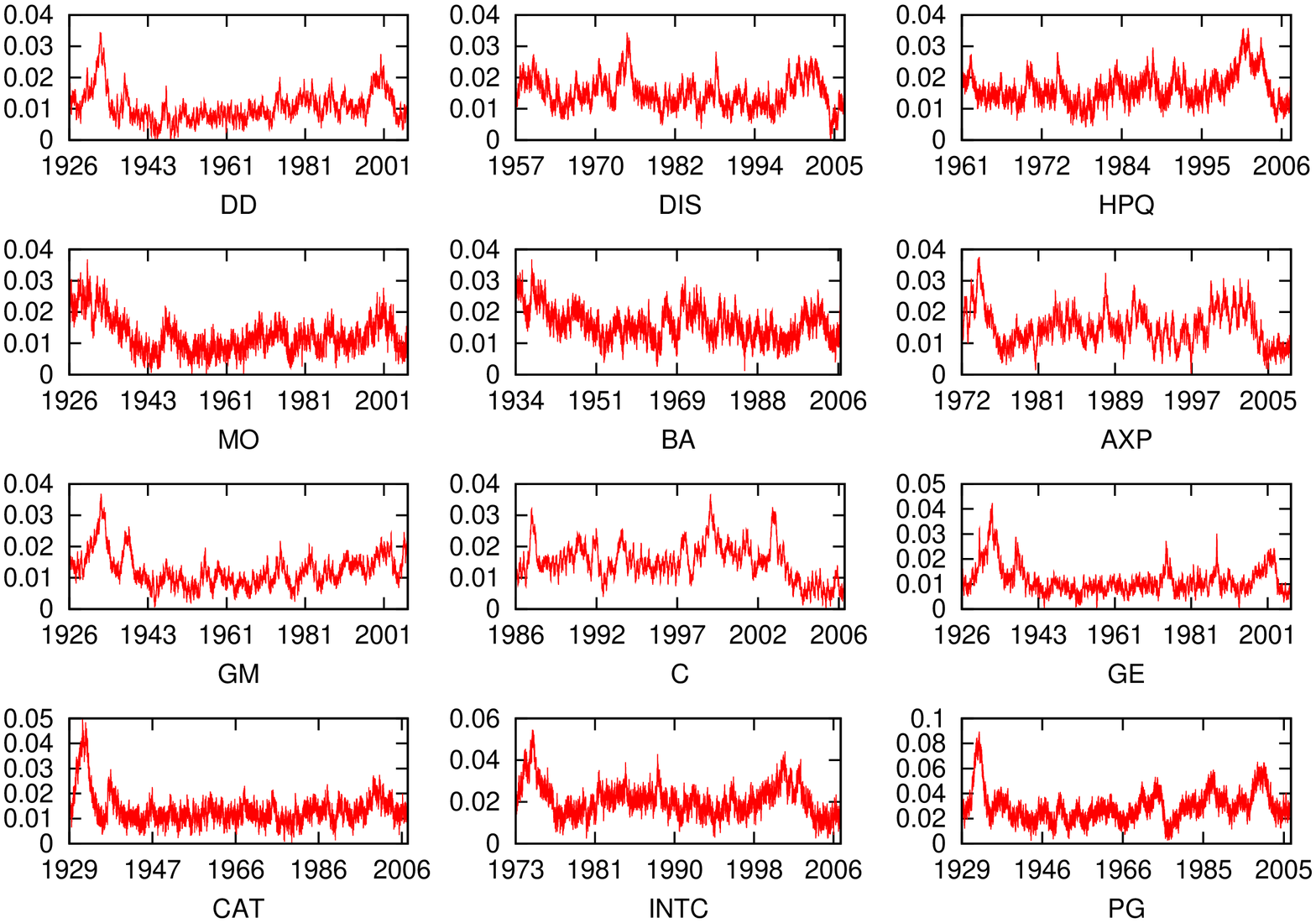}
The Instantaneous Volatility for the DJIA Stocks, II
\end{center}
\caption{\label{Fig-8}
Graphs of the instantaneous volatilities verses trading day for the 12 of
the 24 DJIA stocks with the highest peak volatility are shown.}
\end{figure}

In Figs.~$\ref{Fig-7}$ and $\ref{Fig-8}$, we have graphed the
instantaneous volatility as a function of trading day for all 24
stocks. They have been ordered into graphs where the degree of
volatility are similar, with the stocks with roughly the highest
volatility graphed last. Analytical expressions for the other 23
stocks are not given as they are too lengthy. 

With $\sigma(t)$ now known and the drift for the standardized
daily yield obtained previously, the drift for the daily
yield, $\mu_n\equiv\mu(na)$, can be found for all 24 stocks using the
discretized version of Eq.~$(\ref{muRelation})$, $\mu_n =
\sigma(na)\hat{\mu}_n$, where the $\sigma(t)$ is the instantaneous
volatility obtained above. Since $\vert\hat{\mu}_n\vert < 1$ for all 24
stocks, $\vert\mu_n\vert < \sigma(na)$. Thus for all of the 24 stocks,
the drift of the stock is smaller than the volatility of
it. This is to be expected. If the drift of a stock is larger than the
volatility, then future trends in the stock can be predicted with a
certain degree of certainty; the drift is, after all, a
\textit{deterministic} function of time. Such trends could be seen by
investors, and nearly riskless profits could be made. This clearly does
not happen. It is instead very difficult to discern future trends in
the price of stocks, and this is precisely because the volatility of
the stock is so large.   
 
\section{Concluding Remarks}
\label{sec: 5}

As a continuous process, we have found that the 24 DJIA stocks can be
described as a stochastic process with a volatility that changes
deterministically with time. It is a process for which the
autocorrelation function of the yield vanishes at different times, and
thus one that describes a stock whose price is efficiently
priced. From the results of our calculation of the autocorrelation
function of the daily yield for the 24 stocks, this property of our
stochastic process is in very good agreement with how these stocks are
priced by the market. It is also a process for which the solution of
the stochastic differential can be, at least formally,
solved. This solution is valid only because the volatility is 
a deterministic function of time, however. If the volatility is depends on the
stock price, or if the volatility itself is a stochastic process, the
solution of the stochastic differential equation will not be so
simple, and the autocorrelation function need not vanish at different
times. 

It is, however, only after using the discretized stochastic process
that we are able to 
validate our model. After correcting for the variability of the
volatility by using the standardized daily yield, we have shown
that for all 24 stocks the distribution of standardized daily yields is
well described by the general Rademacher distribution. Indeed, we found
that the abnormally large kurtosis is due to a volatility
that changes with time. For 20 of the 24 stocks, the sample skewness,
kurtosis, and probability distribution agrees with a Rademacher
distribution where $p=1/2$ at the 95\% CL; the probability that these
stocks will increase on any one day is thus equal to the probability
that it will decrease. The other four stocks agree with a generalized
Rademacher distribution and have a $p$ slightly greater than $1/2$. For
these stocks, the probability that the yield will increase on any one
day is slightly higher than the probability that it will decrease. We
conclude that our model is a very good description of the behavior of
these stocks.

That the kurtosis for the standardized daily yield is smaller than the
kurtosis for the daily yield is in agreement with the results found by
Rosenberg (1972). The daily yield is time dependent, and is thus a
nonstationary random variable, while for the standardized daily yield,
the time dependence due to the volatility has been taken account
of. Indeed, in many ways Rosenberg (1972) presages the results of this
work.  

By combining the properties of our continuous stochastic
process for the stocks with noise removal techniques, we have been
able to determine the time dependence of both the volatility and the
drift of all 24 stocks. Unlike the implied volatility, the volatility
obtained here was obtained from the daily close directly without the
need to fit parameters to the market price of options. The theory is
thus self-contained. For Alcoa, Caterpillar, and Johnson\&Johnson, the time
dependence of the volatility can be determined down to a resolution of
a single trading day, while for another 13 stocks, they can be determined to
a resolution of less than 1 1/2 trading days. While other, more
sophisticated signal analysis techniques can be used, given that the
time-series is based on the daily close and thus the resolution is
ultimately limited to a period of two trading days, we do not expect
that it will be possible to dramatically improve on these
results. Only when intraday price data is used will we expect significant 
improvement to this resolution. Indeed, with intraday data we expect
that changes to the volatility that occur during the trading day can be
seen. 

We have deliberately used large cap stocks in our analysis, and
we take care to note that this approach to the analysis of the
temporal behavior of stocks have only been shown to be valid for the
24 stocks we analyzed here. While we would expect it to be applicable
to other large-cap stocks, whether our approach will also be 
valid when applied to mid- or small-cap stocks is still an open
question. Indeed, it will be interesting to see the range of stocks for
which the volatility depends solely on time. 

With both the drift and the volatility determined down nearly to the
single trading day level for most of the stocks, it is now be
possible to calculate the autocorrelation function for both, as
will as the correlation function between the drift and the
volatility. In particular, the degree of influence that the
volatility or drift on any one day has on the volatility or drift on
any future day can be determined. This analysis is currently being
done.



\appendix

\section{Preparing the Time-series}
\label{sec: A1}

The time-series for the 24 DJIA stocks analyzed here were
obtained from the Center for Research in Stock Prices (CRSP). While
the ending date for each series is December 29, 2006, the choice of 
the starting date is often different for different stocks.  This
choice of starting dates was not governed by a desire for uniformity,
but rather by the desire to include as many trading days in the time
series as possible, and thereby minimize standard errors. In addition,
by maximizing the number of trading days included, we also demonstrate
that our model is valid over the entire period for which the prices of
the stock are available.    

Although the daily close of stocks between the starting and ending
dates are used as the basis of the time-series (with dividends
included in the price), a series of adjustments to the CRSP data were made
when the series were constructed. If the closing price of stock is listed by
CRSP as a negative number\textemdash an indication that the closing
price was not available on that day, and the average of the last bid
and ask prices was used instead\textemdash we took the positive
value of this number as the daily close on that day. If no record of
the daily close was given for a particular trading day at
all\textemdash an indication that the bidding and 
asking prices were also not available for that date\textemdash we used the
average of the closing price of the stock on the day preceding and
the day following as the daily close for that day.   

Next, the daily close of the stock prices were scaled to adjust for splits in
the stock. For example, although the daily close for Coca Cola on
December 12, 1925 is listed by CRSP as \$153.625, this 
price was scaled by a factor of 6745.134 to account for the
accumulated splits that the stock has gone through since
1925. The price recorded in the time-series is instead
0.022776. Because of this scale factor, the prices of stocks are
listed in all time-series to an accuracy of at least $10^{-6}$ to
ensure that the daily close on any day can be reconstructed from
the time-series. This level of accuracy or higher was then used in all the
calculations in this paper. While we could have avoided this subtlety
by scaling the daily close, \$48.25, of the stock on December, 29,
2006 by $6745.134$, doing so would result in stock prices that are
\~\$300K, which is deceivingly large. 

Finally, from Eq.~$(\ref{recursion})$ we see that $\sigma_n$ vanishes if
$\Delta u_n$ vanishes, and yet from Eq.~$(\ref{recEq})$, it is
implicit that the quotient $\Delta u_n/\sigma_n$ must be well
defined. Indeed, the reduction of Eq.~$(\ref{stochastic})$ to
Eq.~$(\ref{normal})$ is only valid if $\sigma(t)$ is nowhere zero. In
practice, there are trading days on which the daily yield vanishes;
for Coca Cola, this occurred 2070 out of a total of 
21,523 trading days. To ensure that $\Delta u_n/\sigma_n$ is
well defined on these days, we have added to the daily close a random
number less than $0.00005$ if the close on successive days are
equal. This is done before the daily 
close is scaled to adjust for stock splits. Since the stock price
changes by at least \$0.01 increments, doing so does not materially
change the stock price, while still insuring that $\Delta u_n \ne 0$. 

We have not adjusted for inflation in our time-series, nor have we
accounted for weekends, holidays, or any other days on which 
trading did not take place. We have instead concatenated the daily
close on each trading day, one after another, when constructing the
time-series. The time-series are thus a sequence of \textit{trading} 
days, and not calendar days. While this concatenation is natural,
issues of bias such as those studied by Fleming, Kirby, and Ostdiek
(2006) have not been taken into account. Whether these issues are
relevant for the stocks considered here we leave for further
study. Our focus is instead on the gross features of the stock price. 

Finally, we list here the following particularities that occurred in
our analysis of the 24 DJIA stocks. 

\textit{Boeing:} When solving for $\sigma_n$, the $n=2$ term was
greater than 27, while all other terms was less than $0.5$. This data
point was an outlier, and since it is in the transient
region for the stock, we have set this term equal to $0.1$, which is
the typical size of $\sigma_n$ for $n\ne 2$.

\textit{Merck:} When solving for $\sigma_n$, the $n=2$ term was
greater than 168, while all other terms were three orders of magnitude
smaller. This data point was replaced by $0.2$, which is
the typical size of $\sigma_n$ for $n\ne 2$.

\textit{Exxon-Mobil:} When solving for $\sigma_n$, the $n=2$ term was
greater than 158, while the $n=3$ term was greater than 1500. The
$n=2$ data point was replaced by $0.1$, and the $n=3$ data point was
replaced by $0.009$. 

\section{Statistics}
\label{sec: A2}
In this section, we collect the expressions used here in calculating the
mean, variance, skewness, kurtosis, and autocorrelation of the
time-series, along with their respective standard errors. With the
exception of the autocorrelation function, these expressions are taken
from Stuart and Ord (1994). 

\subsection{Moments and Standard Errors}

Given a collection of $N$ data points, $x_n$, the sample moments,
$m_k$, of order, $k$, that are used in our analysis are defined as
follows   
\begin{eqnarray}
m_1' &\equiv& \frac{1}{N}\sum_{n=1}^N x_n,
\nonumber
\\
m_2 &\equiv& \frac{1}{N-1} \sum_{n=1}^N (x_n-m_1')^2,
\nonumber
\\
m_3 &\equiv& \frac{1}{(N-1)(N-2)}\sum_{n=1}^N(x_n - m_1')^3,
\nonumber
\\
m_4 &\equiv& \frac{N(N+1)}{(N-1)(N-2)(N-3)}\sum_{n=1}^N(x_n - m_1')^4,
\nonumber 
\\
m_5 &\equiv& \frac{N^2(N+5)}{(N-1)(N-2)(N-3)(N-4)}\sum_{n=1}^N(x_n - m_1')^5,
\nonumber 
\\
m_6 &\equiv&
\frac{N(N+1)(N^2+15N-4)}{(N-1)(N-2)(N-3)(N-4)(N-5)}\sum_{n=1}^N(x_n -
m_1')^6, 
\nonumber 
\\
m_8 &\equiv&
\frac{N(N^5+99N^4+757N^3+114N^2-398N+120)}{(N-1)(N-2)(N-3)(N-4)(N-5)(N-6)(N-7)}\sum_{n=1}^N(x_n - m_1')^8,
\label{sample-moments}
\end{eqnarray}
As usual, the sample skewness and kurtosis are defined as
\begin{equation}
skew = \frac{m_3}{m_2^{3/2}}, \qquad kurt = \frac{m_4}{m_2^2}.
\end{equation}

While the standard error of the mean and the variance is well known,
\begin{equation}
\delta m_1' = \sqrt{\frac{m_2}{N}}, \qquad
\delta m_2 = \sqrt{\frac{m_4 - m_2^2}{N}},
\label{SESigma}
\end{equation}
the standard error in the sample skewness and kurtosis are not. For
the skewness, this error is 
\begin{equation}
\delta\, skew =
\frac{1}{\sqrt{N}}\left\{
\frac{m_6}{m_2^3}-6\frac{m_4}{m_2^2} + 9 +
\frac{1}{4}\frac{m_3^2}{m_2^3}\left(9\frac{m_4}{m_2^2}+35\right)
- \frac{3m_5m_3}{m_2^4}
\right\}^{1/2}, 
\label{SESkew}
\end{equation}
while for the sample kurtosis, the standard error is 
\begin{eqnarray}
\delta\, kurt &=&
\frac{1}{\sqrt{N}}\Bigg\{
\frac{m_8}{m_2^4}-4\frac{m_6m_4}{m_2^5}+
4\left(\frac{m_4}{m_2^2}\right)^3
-\left(\frac{m_4}{m_2^2}\right)^2 + 16 \frac{m_4 m_3^2}{m_2^5}
-8\frac{m_5m_3}{m_2^4}+16\frac{m_3^2}{m_2^3}
\Bigg\}^{1/2}. 
\label{SEKurt}
\end{eqnarray}
Although standard errors are defined in terms of the 
population moments, these moments are not known a prior\'i. Following
Stuart and Ord (1994), we have used instead the sample moments listed
in Eq.~$(\ref{sample-moments})$ when calculating standard errors. 

\subsection{The Autocorrelation Function}

For the time-series, $x_n$, where $n = 1, \dots, N$, we define the 
autocorrelation of $x_N$ to be
\begin{eqnarray}
G^{(2)}(x_N, x_{N-M}) \equiv \frac{1}{N-M}\sum_{i=M+1}^{N} &{}&
\left(x_i-\frac{1}{N-M}\sum_{k=M+1}^Nx_{k}\right)
\nonumber
\\
&{}&
\left(x_{i-M}-\frac{1}{N-M}\sum_{k=M+1}^{N}x_{k-M}\right).
\label{GreenFunction} 
\end{eqnarray}
Equation $(\ref{GreenFunction})$ measures the correlation of the
time-series at time step $N$ with the time-series at time step $N-M$. This
definition differs somewhat from the one given in Kendall (1953) and
in Kendall, Stuart and Ord (1983) in that they divide
$G^{(2)}(x_N,x_{N-M})$ by the product of the volatility of the time
series, $\{x_n : n = M+1, \dots, N\}$ with the volatility of
of the time-series, $\{x_{n-M} : n = M+1, \dots, N\}$. It also differs
substantially from the expression used in Alexander (2001), where a
simplified expression for the autocorrelation function in Kendall,
Stuart, and Ord (1983) is used.   

We use Eq.~$(\ref{GreenFunction})$ instead of the expressions given in
Kendall, Stuart, and Ord (1983) and Alexander (2001) for two reasons. First,
$G^{(2)}(x_N,x_N)$ is simply the variance of the time-series, so that the
volatility for a stock can be read off easily from its graph, as can
be seen in Fig.~$\ref{Fig-1}$. Second, we will see below that the
variance for $G^{(2)}(x_N, x_{N-M})$ is easily calculated when $x_n$
are Gaussian random variables, and the standard error for $G^{2}(x_N,
x_{N-M})$ can be readily determined. Derivations of the standard
error for the autocorrelation functions given in Kendall, Stuart, and
Ord (1983) and Alexander (2001), on the other hand, are more involved. 

To determine the standard error for $G^{(2)}(x_N, x_{N-M})$, consider
a time-series where the $x_n$ are Gaussian random variables with mean
zero and standard deviation, $\sigma^2$. Then $E[x_i]=0$ while $E[x_ix_k] =
\sigma^2\delta_{ik}$. (Here, $\delta_{ik}$ is the Kronecker delta with
$\delta_{ik} = 1$ if $i=k$ while $\delta_{ik}=0$ otherwise.) Consequently, 
$E\left[G^{(2)}(x_n, x_{n-M})\right] = 0$ when $M>0$, as can be seen
from Eq.~$(\ref{GreenFunction})$. We thus only have to
calculate  
\begin{equation}
E\left[\left\{G^{(2)}(x_N, x_{N-M})\right\}^2\right]=
\frac{1}{(N-M)^2}\sum_{i=M+1}^N \sum_{k=M+1}^N E[x_ix_{i-M}x_kx_{k-M}].
\end{equation}
Since the $x_n$ are Gaussian random variables,
\begin{equation}
E[x_ix_jx_kx_l] = \sigma^4\left(\delta_{ij}\delta_{kl}+\delta_{ik}\delta_{jl}+\delta_{il}\delta_{jk}\right).
\end{equation}
Thus, 
\begin{eqnarray}
E\left[\left\{G^{(2)}(x_N, x_{N-M})\right\}^2\right] &=& \frac{\sigma^4}{(N-M)^2}\sum_{i,k=M+1}^N
\Big(\delta_{i, i-M}\delta_{k,
  k-M}+\delta_{ik}\delta_{i-M, k-M}+ 
\delta_{i,k-M}\delta_{i-M,k}\Big). 
\nonumber
\\
\label{var}
\end{eqnarray}
The first term vanishes since $M>0$, while the third term vanishes
because it requires that $i = k-M$ and $i-M=k$; this can only happen
when $M=0$. We are thus left with only the second term, so that
\begin{equation}
E\left[\left\{G^{(2)}(x_N, x_{N-M})\right\}^2\right]= \frac{\sigma^4}{N-M}. 
\end{equation}
The standard error, $\Delta G^{(2)}(x_N, x_{N-M})$, for
$G^{(2)}(x_N,x_{N-M})$ when $M>0$ is then simply
\begin{equation}
\Delta G^{(2)}(x_N,x_{N-M}) = \frac{G^{(2)}(x_N, x_N)}{\sqrt{N-M}},
\label{varG}
\end{equation}
where we have used the fact that $G^{(2)}(x_N,x_N)$ is the
variance of the time-series. The standard error for
$G^{(2)}(x_N, x_{N-M})$ when $M=0$ can then be found from
Eq.~$(\ref{SESigma})$ after remembering that $m_4 = 3\sigma^4$ for a
Gaussian distribution. Note that differences between
Eq.~$(\ref{varG})$ and the standard error found in Kendall, Stuart,
and Ord (1983) are due mainly to our defining $G^{(2)}(x_N, x_{N-M})$
with the factor $1/(N-M)$ instead of the factor $1/N$ used by them. 

Although the standard error for $G^{(2)}(x_N, x_{N-M})$ when $x_n$ is
not a Gaussian random variable can be found for special cases (see
Kendall, Stuart, and Ord 1983), Eq.~$(\ref{varG})$ is sufficient for our
purposes. If the market is efficient, we expect the autocorrelation
function for the standard yield to vanish for $M>0$. As
this expectation is borne out by Fig.~$\ref{Fig-1}$, we hypothesize that the
reason why the autocorrelation function in Fig.~$\ref{Fig-1}$ is not
identically zero when $T>0$ is due to sample errors, which in turn is
due to Gaussian random variables with zero mean. We would therefore
expect Eq.~$(\ref{varG})$ to be a good description of the standard
error of this autocorrelation function, and indeed, this expectation
is consistent with the results listed in Table I.

\section{Fourier Analysis}
\label{sec: A3}

In this appendix, we review the properties of the Fourier transform
needed in the analysis we present in this paper. While much of this is
well-known, our purpose here is to establish the notation used in the
paper, and to review the properties of Fourier series needed. At the end of
this section, we will also show that the Fourier transform of a
Gaussian random variable is once again a Gaussian random variable, and
describe the method used to remove the noise from $\sigma_n$.

\subsection{The Discrete Fourier Transform}

Consider a times-series, $x_n$, such that $n = N_{min}, \dots, N_{max}$;
the total number of data points in the time-series is then $N =
N_{max}-N_{min}+1$. In the analysis below, we will assume that
$N$ is an odd number. As  there are at least 5,000 trading days in our
time-series, we can always change the starting point of a time-series
by one trading day to insure that there are an odd number of terms in
the series; this assumption is thus not restrictive.   

The expansion of the time-series in a Fourier series is defined as
\begin{equation}
x_n = \sum_{k=-(N-1)/2}^{(N-1)/2}x_k^\omega e^{-2\pi ikn/N},
\label{FS-x}
\end{equation}
where $i = \sqrt{-1}$. The quantity, $x_k^\omega$, is called the
Fourier transform of $x_n$. As 
\begin{equation}
\exp\left(-\frac{2\pi ikn}{N}\right) = \cos\left(\frac{2\pi
  kn}{N}\right) -  i\sin\left(\frac{2\pi
  kn}{N}\right),
\end{equation}
in taking the Fourier series of $x_n$ we have decomposed $x_n$ into
terms that oscillate with definite period, $T_k = N/k$, for $k>0$, and
have a definite amplitude, $\vert x_k^\omega\vert$. This transform is
thus a natural method of characterizing how a time-series changes with
time.   

The amplitude of these oscillations, $x_k^\omega$, is a complex number
in general. The original time-series, $x_n$, is real, however, and
this fact must also be reflected in $x_k^\omega$. How it is reflected
can be seen by taking the complex conjugate of Eq.~$(\ref{FS-x})$,
\begin{equation}
\bar{x}_n = \sum_{k=-(N-1)/2}^{(N-1)/2}\bar{x}_k^\omega e^{2\pi
  ikn/N},
\label{complex}
\end{equation}
where the complex conjugate is denoted by a bar. Since $x_n =
\bar{x}_n$, by comparing Eq.~$(\ref{FS-x})$ with
Eq.~$(\ref{complex})$ we find after taking $k\to-k$ in
Eq.~$(\ref{FS-x})$ the reality condition
$\bar{x}_k^\omega=x_{-k}^\omega$ that the Fourier transform must
satisfy.

The transform Eq.~$(\ref{FS-x})$ is invertible. Namely, we can express 
$x_k^\omega$ in terms of $x_n$ by taking the following sum
\begin{equation}
\sum_{n=N_{min}}^{N_{max}}x_n \exp\left(\frac{2\pi\hat{k}n}{N}\right)=
\sum_{k=-(N-1)/2}^{(N-1)/2} x_k^\omega\sum_{n=N_{min}}^{N_{max}}
\left[\exp\left(\frac{2\pi i(\hat{k}-k)}{N}\right)\right]^n,
\label{inverse}
\end{equation}
of Eq.~$(\ref{FS-x})$. To evaluate this sum, we consider first the case
where $\hat{k}-k \ne qN$ for any integer $q$. The series on the right
can then be summed to give 
\begin{equation}
\sum_{n=N_{min}}^{N_{max}}
\left(e^{2\pi i(\hat{k}-k)/N}\right)^n =
e^{\left[2\pi i(\hat{k}-k)N_{min}/N\right]}\left(\frac{1-e^{2\pi
    i(\hat{k}-k)}}{1-e^{2\pi i(\hat{k}-k)/N}}\right),
\label{sum2}
\end{equation}
after using the following identity for the geometric series, 
\begin{equation}
\sum_{n=0}^Ny^n = \frac{1-y^{N+1}}{1-y}.
\end{equation}
As $e^{2\pi i(\hat{k}-k)}=1$, while $e^{2\pi
  i(\hat{k}-k)/N}\ne1$, we conclude that Eq.~$(\ref{sum2})$
vanishes in this case. We next consider the case when $\hat{k}-k =
qN$. Each term in the sum is then one, and Eq.~$(\ref{sum2})$ is 
easily summed to give $N$. 

Combining these two results, we find that 
\begin{equation}
\sum_{n=N_{min}}^{N_{max}}
\left(e^{2\pi i(\hat{k}-k)/N}\right)^n = N\delta_{\hat{k},k}.
\label{sum}
\end{equation}
We then conclude from Eq.~$(\ref{inverse})$ that
\begin{equation}
x_k^\omega =\frac{1}{N} \sum_{n=N_{min}}^{N_{max}}x_n e^{2\pi ikn/N}.
\end{equation}
This is the inverse Fourier transform of $x_n$. In particular, notice that when
$k=0$, 
\begin{equation} 
x_0^\omega =
\frac{1}{N}\sum_{n=N_{min}}^{N_{ max}}x_n,
\end{equation}
is simply the average of $x_n$ over the whole time-series.

The Fourier series Eq.~$(\ref{FS-x})$ can also be expressed as an
explicitly real expansion, 
\begin{equation}
x_n = a_0^{\cos} +2\sum_{k=1}^{(N-1)/2}a_k^{\cos}\cos\left(\frac{2\pi
  kn}{N}\right) + 2\sum_{k=1}^{(N-1)/2}a_k^{\sin}\sin\left(\frac{2\pi
  kn}{N}\right), 
\label{real}
\end{equation}
where the amplitudes
\begin{eqnarray} 
a_k^{\cos} &\equiv& \frac{1}{2}\left(x_k^\omega + \bar{x}_k^\omega\right)
= \frac{1}{N}\sum_{n=N_{min}}^{N_{max}} x_n \cos(2\pi nk/N),
\nonumber
\\
a_k^{\sin} &\equiv& \frac{1}{2i}\left(x_k^\omega -
\bar{x}_k^\omega\right)
= \frac{1}{N}\sum_{n=N_{min}}^{N_{max}} x_n \sin(2\pi nk/N).
\label{aSinCos}
\end{eqnarray}
are called the Fourier cosine and Fourier sine coefficients,
respectively. While analytical calculations are more easily done with
Eq.~$(\ref{FS-x})$, numerical calculations are necessarily done with
Eq.~$(\ref{real})$, and it is on the Fourier sine and cosine
coefficients that we will focus most of our analysis in this paper.

\subsection{Fourier Transforms of Gaussian Random Variables}

We now prove the theorem stated in Sec.~$\ref{sec: 4}$ for the special case
when $N$ is an odd number. Although the theorem holds in general, this
is the only case we need here.

Because each $\xi_n$ in the time-series is a Gaussian random
variable with zero mean and variance, $\sigma^2$, the probability
distribution for the time-series is just
\begin{equation}
P(\xi_1, \dots, \xi_N)
=\frac{1}{\sigma^N}\left(\frac{a}{2\pi}\right)^{N/2}
\prod_{n=1}^Ne^{-\xi_n^2/2\sigma^2} = 
\frac{1}{\sigma^N}\left(\frac{a}{2\pi}\right)^{N/2}
\exp\left(-\frac{1}{2\sigma^2}\sum_{n=1}^N \xi_n^2\right),
\label{Prob}
\end{equation} 
where $a$ is the time interval between successive points in the time
series, and in the last equality we have used $E[\xi_n\xi_m] = 0$ for $n\ne
m$. Expanding $\xi_n$ in a Fourier series using Eq.~$(\ref{FS-x})$, we
find that  
\begin{equation}
\sum_{n=1}^{N}\xi_n^2 = \sum_{k, k'=-(N-1)/2}^{(N-1)/2}\xi^\omega_k
\xi^\omega_{k'}\sum_{n=1}^N \exp\left[-\frac{2\pi in}{N}(k+k')\right]
= \sum_{k, k'=-(N-1)/2}^{(N-1)/2}\xi^\omega_k
\xi^\omega_{k'}N \delta_{k, -k'},
\label{KronDelta}
\end{equation}
where the last equality holds from Eq.~$(\ref{sum})$. Thus,
\begin{equation}
\sum_{n=1}^{N}\xi_n^2 =
N\sum_{k=-(N-1)/2}^{(N-1)/2}\vert\xi^\omega_k\vert^2,
\label{amp}
\end{equation}
which is Parseval's Theorem for a discrete Fourier series. Following
Eq.~$(\ref{aSinCos})$, we express
$\vert\xi_k^\omega\vert^2=\left(\alpha^{\sin}_k\right)^2
+\left(\alpha^{\cos}_k\right)^2$ in Eq.~$(\ref{amp})$.  Then 
Eq.~$(\ref{Prob})$ can be written as 
\begin{eqnarray}
P=\frac{1}{\sigma^N}\left(\frac{a}{2\pi}\right)^{N/2}
\prod_{k=-(N-1)/2}^{(N-1)/2}
  \exp\left\{-\frac{N}{2\sigma^2}\left[\left(\alpha^{\sin}_k\right)^2+\left(\alpha^{\cos}_k\right)^2\right]\right\},  
\label{FProb}
\end{eqnarray}
and the theorem is proved.

It is straightforward to see that the converse is also
true. Namely, if $\alpha_k^{\cos}$ and $\alpha_k^{\sin}$ are Gaussian random
variables with zero mean and volatility $\sigma/\sqrt{N}$, then $x_n$
is a Gaussian random variable with zero mean and volatility, $\sigma$.

Notice from Eq.~$(\ref{FProb})$ that $E[\alpha^{\cos}_k \alpha^{\sin}_k] =0$,
and thus the two random variables are independent. Notice also that while
we began with $N$ degrees of freedom with the random variables, $x_n$,
we seem to have ended up with $2N-1$ degrees of freedom for
the random variables, $\alpha^{\cos}_k$ and $\alpha^{sin}_k$. From
Eq.~$(\ref{aSinCos})$ we see, however, that $\alpha_k^{\cos} =
\alpha_{-k}^{\cos}$ and $\alpha_k^{\sin} = \alpha_{-k}^{\sin}$; not
all the variables in Eq.~$(\ref{FProb})$ are independent. When this
redundancy is taken account of, we arrive back to $N$ degrees of freedom.  

\subsection{Removal of Noise}
\label{sec: A3-noise}

In this subsection, we will describe how the noise present in $\sigma_n$
is removed, and how $\sigma(t)$ is obtained.  

Given that the noise floor associated with the Fourier sine and cosine
coefficients is constant over all frequencies, $f_k$, noise removal is
straight forward. We need only remove from $\mathcal{F}^{\sin}$, the
set of all Fourier sine coefficients for $\sigma_n$, and  
$\mathcal{F}^{\cos}$, the set of all Fourier cosine
coefficients for $\sigma_n$, those coefficients whose amplitudes is
less than the amplitudes, $S_{noise}$ and $C_{noise}$, of the noise
floor for the Fourier sine and Fourier cosine coefficients,
respectively. The coefficients left over\textemdash
$\mathcal{F}^{\sin}_{signal} = 
\{a^{\sin}_k \in \mathcal{F}^{\sin}: \vert a^{\sin}_k \vert >
S_{noise}\}$  for the Fourier sine coefficients, and
$\mathcal{F}^{\cos}_{signal} = \{a^{\cos}_k \in \mathcal{F}^{\cos} :
\vert a^{\cos}_k \vert > C_{noise}\}$ for the Fourier cosine
coefficients\textemdash can then be used to construct the instantaneous
volatility, $\sigma(t)$, by summing the Fourier series Eq.~$(\ref{signal})$.  

The noise floor amplitudes, $S_{noise}$ and $C_{noise}$, are determined
statistically. Consider the set of coefficients that are removed:
$\mathcal{F}^{\sin}_{noise}= \{a^{\sin}_k\in \mathcal{F}^{\sin} :
\vert a^{\sin}_k \vert \le S_{noise}\}$ for the Fourier sine
coefficients and $\mathcal{F}^{\cos}_{noise}= \{a^{\cos}_k\in 
\mathcal{F}^{\cos} : \vert a^{\cos}_k \vert \le
C_{noise}\}$ for the Fourier cosine coefficients. Because the
distribution of the noise floor is Gaussian, $S_{noise}$ and
$C_{noise}$ must be chosen so that the distributions of
coefficients in $\mathcal{F}^{\sin}_{noise}$ and
$\mathcal{F}^{\cos}_{noise}$ are Gaussian as well. If either
amplitude is chosen too large, then coefficients from 
$\mathcal{F}^{\sin}$ or $\mathcal{F}^{\cos}$ 
that make up the signal, $\sigma(t)$, would be included in the noise
distributions as noise. As these coefficients are 
supposed to be above the noise, they will skew and flatten the
distribution; the skewness and the kurtosis for the distribution of 
$\mathcal{F}^{\sin}_{noise}$ and of $\mathcal{F}^{\cos}_{noise}$ will
then differ from their Gaussian values if these coefficients are
included. On the other hand, if either amplitude for the noise floor
is chosen too \textit{small}, then coefficients from
$\mathcal{F}^{\sin}$ or $\mathcal{F}^{\cos}$ that make up the noise
would be \textit{excluded} from the noise distributions. As these
coefficients would have populated the tails of the Gaussian
distribution, their removal will tend to \textit{narrow} the
distribution, and the kurtosis of the noise distributions will differ
once again from its Gaussian value. (Because the coefficients are
remove symmetrically about the horizontal zero line, a choice of the
amplitude for the noise floor that is too small will not tend to
change the skewness significantly.) Thus, $S_{noise}$  and $C_{noise}$
must be chosen so that the skewness and kurtosis of the distribution
of coefficients in $\mathcal{F}^{\sin}_{noise}$ and
$\mathcal{F}^{\cos}_{noise}$ is as close to their Gaussian
distribution values as possible. 

While the above procedure is straightforward, there is an additional
constraint. The volatility cannot be negative, and thus the
resultant instantaneous volatility, $\sigma(t)$, obtained after the
noise floor is removed must the positive as well. This constraint is
not trivial. For a number of stocks, a choice of $S_{noise}$ and
$C_{noise}$ that results in noise distributions that are closest to a
Gaussian distribution also results in a $\sigma(t)$ that is negative
on certain days. To obtain a $\sigma(t)$ that is non-negative,
slightly larger amplitudes for the noise floors were chosen, which
resulted in a slightly larger skewness and kurtosis.

This approach to removing the noise from the volatility,
$\sigma_n$, has been successfully applied to all 24 stocks using a
simple C++ program that implements an iterative search algorithm to
determine $S_{noise}$ and $C_{noise}$. The results of our numerical
analysis are shown in Table IV. There, we have listed the noise floor
amplitudes, $S_{noise}$ and $C_{noise}$, used for each of the 24
stocks. Their values are given as multiples of the standard deviation
of the distribution of the Fourier coefficients in
$\mathcal{F}^{\sin}_{noise}$ and $\mathcal{F}^{\cos}_{noise}$. As
these values range from 3.030 times the standard deviation to 3.948
times the standard deviation, 99.756\% to 99.992\% of the data points
that make up a Gaussian distribution can be included in these
distributions if they are present in either
$\mathcal{F}^{\sin}_{noise}$ or $\mathcal{F}^{\cos}_{noise}$.   

Listed also in Table IV are the kurtosis for
$\mathcal{F}^{\sin}_{noise}$ and $\mathcal{F}^{\cos}_{noise}$. We have
found that they range in value from 2.95 to 3.31, and are thus very
close to the Gaussian distribution value of three for the
kurtosis. The skewness of the noise of the distribution of
$\mathcal{F}^{\cos}_{noise}$ was calculated as well, and was found to
vary in value from -0.03 to 0.11; this also is very close to the
Gaussian distribution value of zero for the skewness. The skewness for
the distribution of $\mathcal{F}^{\sin}_{noise}$ was also calculated,
but we find that their values are $10^{-2}$ to $10^{-7}$ times smaller
than the skewness for the distribution of
$\mathcal{F}^{\cos}_{noise}$, and there was no need to listed these
values in the table. This extremely close agreement with the  skewness
of the Gaussian distribution is because the Fourier sine coefficients
are antisymmetric about $k=0$: $a_k^{\sin}=-a^{\sin}_{-k}$. The
average of any odd power of $a^{\sin}_k$ over $k$\textemdash and in
particular, the skewness of the distribution of
$\mathcal{F}^{\sin}$\textemdash thus automatically vanishes. For this
reason, the skewness for $\mathcal{F}^{\sin}_{noise}$ is exceedingly small.  

\begin{acknowledgments}

The author is also an adjunct professor at the Department of Physics,
Diablo Valley College, Pleasant Hill, CA 94523, and a visiting
professor at the Department of Physics, University of California,
Berkeley, CA 94720. He would like to thank Peter Sendler for his support and 
helpful criticisms while this research was being done. It is doubtful
that this work would have been completed without his encouragement
and, yes, prodding.

\end{acknowledgments}

\end{document}